\documentclass{aa}  

\usepackage{graphicx}
\usepackage{multirow}
\usepackage{xcolor}
\usepackage{hyperref}
\hypersetup{
    colorlinks = true,
    citecolor = {.},
    linkcolor = {red},
    urlcolor = {blue}
}

\newcommand{\bxi}{\boldsymbol{\xi}}

\usepackage{natbib}
\bibpunct{(}{)}{;}{a}{}{,} 

\usepackage{txfonts}

\makeatletter
\renewcommand*\aa@pageof{, page \thepage{} of \pageref*{LastPage}}
\makeatother

\begin{document}

   \title{Acoustic wave propagation through solar granulation:\\
   Validity of effective-medium theories, coda waves}
   \titlerunning{Acoustic wave propagation through solar granulation}

   \author{P.-L. Poulier
          \inst{1}
          \and
          D. Fournier\inst{1}
          \and L. Gizon\inst{1,2}
          \and T. L. Duvall Jr.\inst{1}
          }

   \institute{Max Planck Institute for Solar System Research, Justus-von-Liebig-Weg 3, 37077 G{\"o}ttingen,
    Germany \\
    \email{poulier@mps.mpg.de}
              \and
              Georg-August-Universit{\"a}t, Friedrich-Hund-Platz 1, 37077 G{\"o}ttingen, Germany\\
             }

   \date{Received ; accepted }
 
  \abstract
   {The frequencies, lifetimes, and eigenfunctions of solar acoustic waves are affected by turbulent convection, which is random in space and in time. 
   Since the correlation time of solar granulation and the periods of acoustic waves ($\sim$5~min) are similar, the medium in which the waves propagate cannot a priori be assumed to be time independent.
   }
   {
   We compare various effective-medium solutions with numerical solutions in order to identify the approximations that can be used in helioseismology.
   For the sake of simplicity, the medium is one dimensional.
   }
   {
   We consider the Keller approximation, the second-order Born approximation, and spatial homogenization to obtain theoretical values for the effective wave speed and attenuation (averaged over the realizations of the medium).
   Numerically, we computed the first and second statistical moments of the wave field over many thousands of realizations of the medium (finite-amplitude sound-speed perturbations are limited to a 30 Mm band and have a zero mean).}
   {
   The effective wave speed is reduced for both the theories and the simulations.
   The attenuation of the coherent wave field and the wave speed are best described by the Keller theory.
   The numerical simulations reveal the presence of coda waves, trailing the coherent wave packet. These late arrival waves are due to multiple scattering and are easily seen in the second moment of the wave field.
   }
   {
   We find that the effective wave speed can be calculated, numerically and theoretically, using a single snapshot of the random medium (frozen medium); however, the attenuation is underestimated in the frozen medium compared to the time-dependent medium.
   Multiple scattering cannot be ignored when modeling acoustic wave propagation through solar granulation.
  }


   \maketitle
%


\section{Introduction}

Solar seismic waves interact with small-scale convective motions near the solar surface via a wave scattering process, affecting their properties (e.g., propagation speed, frequency, amplitude, and phase). As the e-folding lifetime of  solar granulation is comparable to the period of the waves, the medium may not be assumed to be frozen. Furthermore, the spatial spectrum of convection encompasses all scales, including those that are comparable to the wavelengths of p and f modes.

Most approaches that have been proposed so far assume a separation of scales between the waves and the medium. Often the wave period is assumed to be much smaller than the time scale of the evolution of convective flows \citep{1984Sci...226..687B,Delache1988,1999A&A...351..689R}. \cite{1993A&A...272..595M, 1993A&A...272..601M}, using this assumption, derived a model for the scattering of the f~mode by granulation using the binary collision approximation \citep{Howe1971} and found a mode frequency reduction due to the scattering as well as a large attenuation, which compare favorably with observations \citep{1998ApJ...505L..55D}. Other authors assume that the wave period, or the wavelength, is much larger than the temporal, or the spatial, scale of convection, which allows one to apply homogenization techniques \citep{2013ApJ...773..101H,2015ApJ...806..246B}. 

Numerical simulations provide a useful means to study the interaction of seismic waves with convection \citep[e.g.,][]{2016A&A...592A.159B,2017MNRAS.464L.124H,Schou2020}. Turbulent convection has an indirect effect on the waves through a change in the average medium  (e.g., via a turbulent pressure term) and, in addition, it affects the physics of wave propagation and attenuation via a scattering process.

Here, we study this problem under a highly simplified setup. We consider a one-dimensional steady medium that contains sound-speed perturbations over a finite region, but other than that it is uniform.
There is no a priori separation of scales in space nor in time between the incoming  wave packet and the medium. For relative sound-speed perturbations of a significant amplitude (e.g., 5\% and above) multiple scattering plays a significant role in the redistribution of wave energy. We compare our numerical simulations with theoretical approximations, which are easy to implement in this context.

Since the medium is random in both time and space, we study the effect of the medium on the waves in a statistical sense by computing the first and second moments of the quantities of interest (e.g., the wave field) over many realizations. From the expectation value of the wave field, also known as the coherent or ballistic wave field, we can extract the 
attenuation and the effective wave speed for example. For the variance of the wave field, we can extract information about the distribution of backward- and forward-scattered energy. This includes late-arrival fluctuations due to multiply-scattered (coda) waves. 

 We state the problem in Section~\ref{Model_section} and explain the numerical implementation in Section~\ref{sect:numerics}. Various effective medium theories are reviewed in Section \ref{sect:theories}. We present our results in Section~\ref{Results_section}, and discuss them in Section~\ref{sect:discussion}.

\section{Statement of the problem \label{Model_section}}

\subsection{The random medium} \label{sect:random}

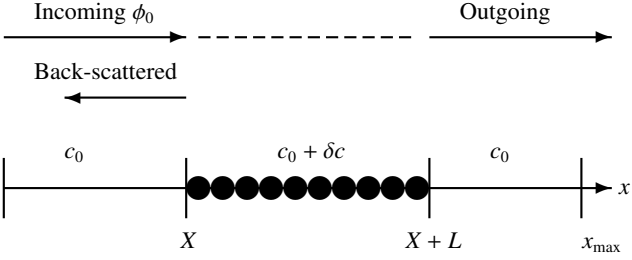
\begin{figure}[ht]

\setlength{\unitlength}{0.8cm}
\begin{picture}(10.5,6)

\thicklines

\put(0,2){\vector(1,0){10}}
\put(10.1,1.9){{\footnotesize $x$}}
\put(0,1.5){\line(0,1){1}}
\put(9.5,1.5){\line(0,1){1}}
\put(9.5,1){{\footnotesize $x_{\rm max}$}}

\put(3,1.5){\line(0,1){1}}
\put(2.9,1){{\footnotesize $X$}}
\put(7,1.5){\line(0,1){1}}
\put(6.6,1){{\footnotesize $X+L$}}

\put(3.2,2){\circle*{0.4}}
\put(3.6,2){\circle*{0.4}}
\put(4,2){\circle*{0.4}}
\put(4.4,2){\circle*{0.4}}
\put(4.8,2){\circle*{0.4}}
\put(5.2,2){\circle*{0.4}}
\put(5.6,2){\circle*{0.4}}
\put(6,2){\circle*{0.4}}
\put(6.4,2){\circle*{0.4}}
\put(6.8,2){\circle*{0.4}}

\put(1,2.5){{\footnotesize $c_0$}}
\put(4.5,2.5){{\footnotesize $c_0+\delta c$}}
\put(8,2.5){{\footnotesize $c_0$}}

\put(0,4.5){\vector(1,0){3}} 
\put(0.5,4.8){{\footnotesize Incoming $\phi_0$}}
\put(3,3.5){\vector(-1,0){2}} 
\put(0.5,3.8){{\footnotesize Back-scattered}}
\put(7,4.5){\vector(1,0){3}} 
\put(7.5,4.8){{\footnotesize Outgoing}}
\multiput(3.2,4.5)(0.3,0){12}{\line(1,0){0.2}}

\end{picture}

\caption{Schematics  of the problem.}
\label{pic:geometry}

\end{figure}

We consider a uniform one-dimensional background with sound speed $c_0=10 \ {\rm km/s}$, a value of the same order of magnitude as the sound speed at the solar surface. We perturb the medium by adding locally a space- and time-dependent random fluctuation:
\begin{equation}
  c(x,t)=\begin{cases}
               c_0 &  (x< X) \\
               c_0+\delta c(x,t) &  (X \le x \le X+L) \\
               c_0 &  (x>X+L). \\
            \end{cases}
\label{sound_speed_perturbation}
\end{equation}
This is shown in Fig.~\ref{pic:geometry}, where the filled circles symbolize the fluctuation. In Eq.~\eqref{sound_speed_perturbation}, $\delta c$ has a zero mean so that $\langle c \rangle = c_0$ where angle brackets   denote an expectation value.

\begin{figure}[ht]
    \centering
    \includegraphics[width=0.5\textwidth]{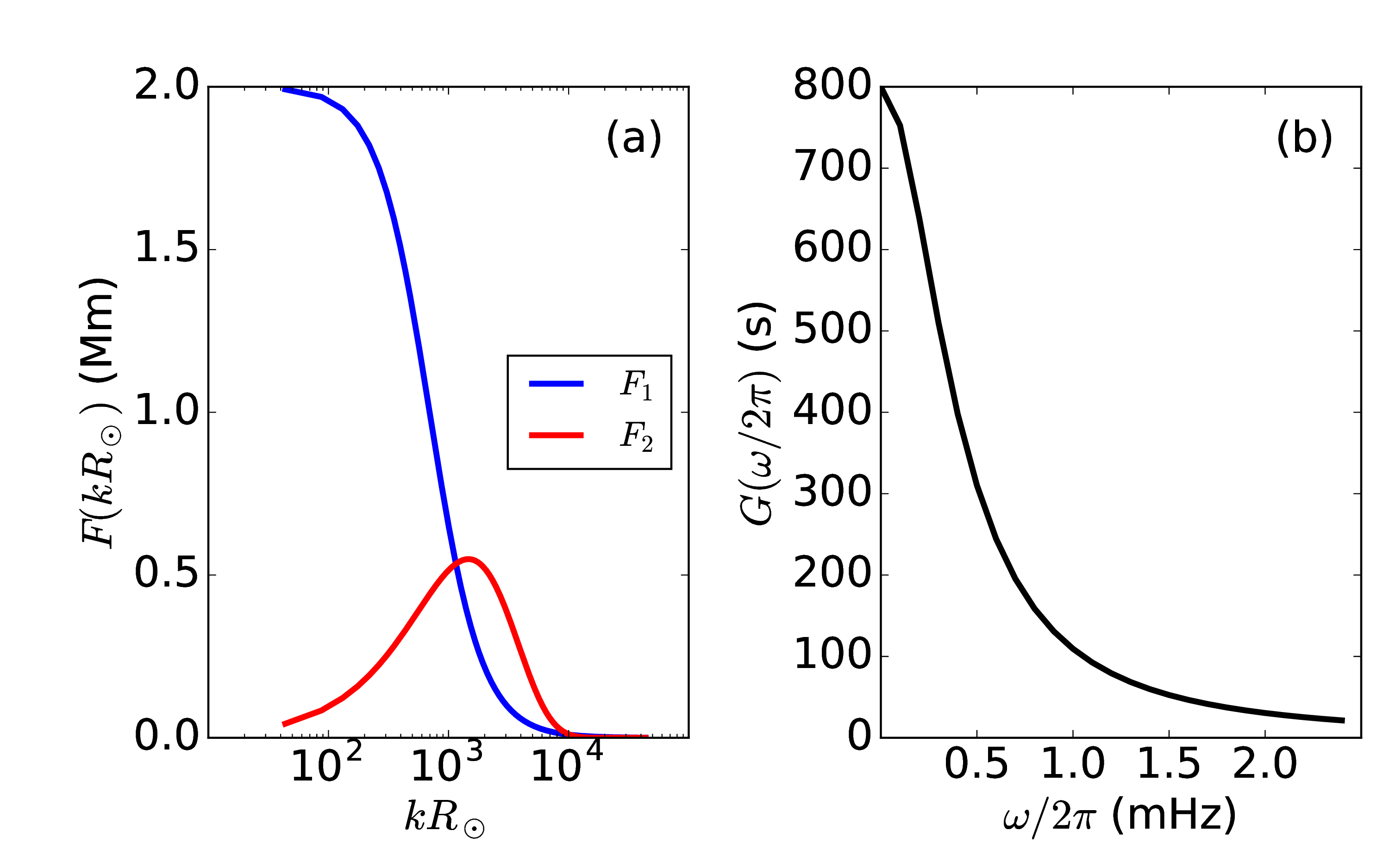}
    \includegraphics[width=0.5\textwidth]{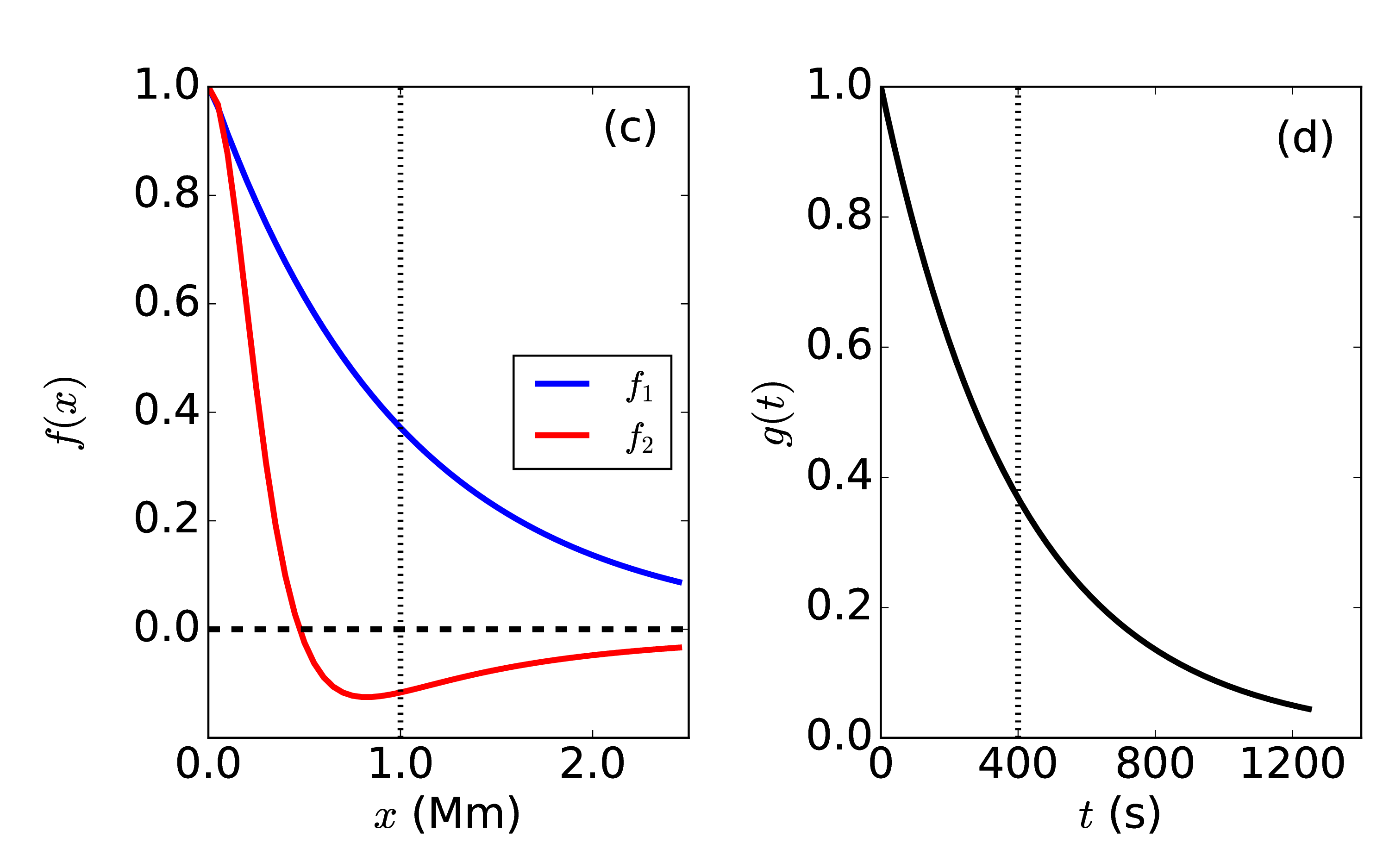}
    \caption{(a): Power spectrum as a function of the adimensional wave number $k R_\odot$. (b): Power spectrum as a function of frequency. (c): Spatial autocorrelation. (d): Temporal autocorrelation. The vertical dotted lines are drawn at the values of the correlation parameters chosen for medium 1, namely $\tau=400$~s and $a=1$~Mm.}
    \label{power_spectrum_medium}
\end{figure}

The sound-speed perturbation is specified through the autocorrelation
\begin{equation}
\langle \delta c  (x',t')\ \delta c (x'+x,t'+t) \rangle =  \epsilon^2 c_0^2 f(x) g(t),
\label{eq:autocor}
\end{equation}
where we assume a separation between time and space. The value of $\epsilon$ is at most $0.1$ in our simulations. The random medium can equivalently be characterized by its power spectrum 
\begin{equation}
    P(k,\omega) = \int   f(x) e^{-i kx} \mathrm{d}x
    \int 
    g(t) e^{ i\omega t } \mathrm{d}t =  F(k) G(\omega) .
\end{equation}
In time, we choose an exponential profile
\begin{equation}
    g(t) = \exp(-|t|/\tau),
\end{equation}
where $\tau$ is the e-folding lifetime. For granulation, we have $\tau \approx 400 \ {\rm s}$  \citep[e.g.,][]{Title1989}. The temporal power spectrum is  Lorentzian,
\begin{equation}
    G(\omega) = \frac{2\tau}{1+(\omega\tau)^2}.
\end{equation}
In space, we consider two different types of profile. The first choice is an exponential medium (hereafter Medium 1), which will enable us to carry out approximations  analytically:
    \begin{equation}
        f_1(x) =  \exp(-|x|/a).
    \end{equation}
For a granulation-like medium, it is reasonable to choose     $a=1 \ {\rm Mm}$. In Fourier space,
    \begin{equation}
        F_1(k) = \frac{2a}{1+(ka)^2}.
    \end{equation}
The second choice (hereafter Medium 2) is a spatial power spectrum of the form \citep[e.g.,][]{2013AASP....3...89B}
    \begin{equation}
        F_2(k) = C |k|^\alpha \exp(-\beta |k|)  ,
    \label{eq:power_granulation}
    \end{equation} 
    where $C = \pi\beta^{\alpha+1}/\Gamma(\alpha+1)$ is a normalization factor such that the spatial autocorrelation function equals $1$ at $x=0$.
    The parameters $\alpha$ and $\beta$ can be tuned to obtain a power spectrum that peaks at the desired wavenumber.
   Here we fix $\alpha=1$ and $\beta = 6.7\times 10^{-4} R_\odot$ where $R_\odot=696$~Mm, such that the spatial power spectrum peaks at $k R_\odot=1500$.
    In real space, for $\alpha=1$, we have
    \begin{equation}
       f_2(x) = \frac{1-(x/\beta)^2}{(1+(x/\beta)^2)^2} .
    \end{equation}
The two power spectra and their corresponding autocorrelation functions are shown in Fig.~\ref{power_spectrum_medium}.

From the knowledge of the power spectrum $P(k,\omega)$, we can compute a realization of the sound speed perturbations as follows:
\begin{equation}
    \delta c(x,t) = \frac{c_0}{(2\pi)^2} 
    \int
    \sqrt{P(k,\omega)} \mathcal{N}(k,\omega)\ e^{i (kx - \omega t)}\  \mathrm{d}\omega \mathrm{d}k ,
\end{equation} 
where $\mathcal{N}(\omega,k)$ is realization of a complex Gaussian random variable with zero mean and unit variance (the real and the imaginary parts are independent). To ensure that $\delta c(x,t)$ is real, we have $\mathcal{N}(k,\omega)=\mathcal{N}^*(-k,-\omega)$. This way to proceed is based on the assumptions of stationarity and horizontal spatial homogeneity of the medium 
\citep[e.g.,][]{Gizon2004}.

\subsection{The wave equation}

The displacement $\bxi$ of acoustic waves is given by \citep{1967MNRAS.136..293L}
\begin{equation}
    \partial_t^2 \bxi - \frac{1}{\rho} \nabla \left( \rho c^2 \nabla \cdot \bxi \right) = 0.  \label{eq:vectorial}
\end{equation}
Here, we have ignored gravity, rotation, damping as well as any background flows. This equation has been derived in a background medium where the parameters $\rho$ and $c$ are independent of time. However, \citet{legendre2003these} showed that this formulation remains valid for a time-varying medium. Taking the divergence of Eq.~\eqref{eq:vectorial} and denoting $\phi = \nabla \cdot \bxi$, we obtain
\begin{equation}
    \partial_t^2 \phi - \nabla \cdot \left( \frac{1}{\rho} \nabla (\rho c^2 \phi) \right) = 0. 
\end{equation}
In this paper, we assume that the density is constant and consider the following 1D acoustic wave equation
\begin{equation}
\partial_t^2 \phi - \partial_x^2 (c^2 \phi) = 0.
\label{bible_equation}
\end{equation}
We implement two numerical codes. The first code is a  time-domain code to study the propagation of the wavepacket through a time-dependent random medium, based on Eq.~\eqref{bible_equation}. As  initial condition, we inject at  location $x_0<X$ a wave packet of central frequency $\omega_0$ and frequency width $\sigma$:
\begin{equation}
    \phi(x,0) = \phi_0(x,0), \quad \partial_t \phi(x,0) = \partial_t \phi_0 (x,0),
\end{equation}
where
\begin{equation}
    \phi_0(x,t) = \exp \left[ -\frac{\sigma^2}{2}\left(\frac{x-x_0}{c_0}-t\right)^2 \right]\cos \left[ \omega_0 \left(\frac{x-x_0}{c_0}-t\right) \right].
\end{equation}
As shown in the schematics of Fig.~\ref{pic:geometry}, the incoming wave packet first travels in the $+x$ direction in the homogeneous medium, experiences scattering inside the perturbed medium, then comes out (outgoing wave packet) and propagates in the $+x$ direction in the homogeneous medium. Part of the wave packet is back-scattered and travels in the $-x$ direction. The simulation box is large enough so that the wave packet is not affected by the  computational boundaries at $x=0$ and $x=x_{\rm max}$.
    
The second code is a frequency-domain code to study the  wave field in a frozen medium ($\tau \rightarrow \infty$). For a sound speed that does not depend on  time, we can take the temporal Fourier transform of Eq.~\eqref{bible_equation} to obtain the wave equation in the frequency domain, i.e. the Helmholtz equation
\begin{equation}
    \partial_x^2 (c^2\tilde \phi(x,\omega)) + \omega^2 \tilde \phi(x,\omega) = 0, \label{eq:helmholtz}
\end{equation}
with Dirichlet boundary condition at $x=0$,
\begin{equation}
    \tilde \phi(0, \omega)=1,
\end{equation}
and the Sommerfeld outgoing radiation condition
\begin{equation}
    \partial_x \tilde \phi(x_{\rm max}, \omega)= \frac{ i \omega}{c(x_{\rm max})}\tilde \phi(x_{\rm max}, \omega).
\end{equation}
The tilde denotes the temporal Fourier transform.

\subsection{Characterizing the wave field}

The wave field is affected randomly by the perturbations. The statistical effects can however be studied by looking at the moments of the wave field, i.e. by doing some averages over the realizations of the random medium.

In particular, the coherent wave field is attenuated because each wave packet travels in a different random realization of the medium and is deformed in a different way. This damping is related to the lifetime of the average acoustic wave. The coherent wave field also propagates with a different velocity than $c_0$, depending on frequency, called the effective wave speed.

An approximate representation of the coherent wave field inside the perturbed medium is therefore
\begin{equation}
    \langle \psi \rangle \sim e^{  i k(\omega) x - i \omega t } = e^{-k_i(\omega) x} e^{i  k_r(\omega) x - i \omega t } ,
    \label{representation}
\end{equation}
where the effective wave number is
\begin{equation}
k_r(\omega) = {\rm Re}\ k(\omega), 
\end{equation}
and the spatial attenuation is 
\begin{equation}
k_i (\omega) = {\rm Im}\ k(\omega) .
\end{equation}
The effective wave speed is defined by
\begin{equation}
     c_{\rm eff}(\omega) = \frac{\omega}{ k_r(\omega)}.
\end{equation}
Similarly, we define $k_0(\omega)$, the wave number for an unperturbed wave field, such that
\begin{equation}
    c_0 = \frac{\omega}{k_0(\omega)}.
\end{equation}
We want to solve the (simplified) problem of acoustic wave scattering numerically and find which approximations work to retrieve the coherent wave field. In particular, we check whether we can get rid of the time dependence and assume a frozen medium.

Furthermore, we want to investigate the phenomenon of multiple scattering due to the finite-amplitude perturbations. This is more easily done by looking at the second moment (variance) of the wave field. It contains information that is otherwise zeroed out by doing a mere average. By doing so, the coda waves, which trail the ballistic wave packet and are often studied in seismology, can be readily observed.

\section{Numerical methods} \label{sect:numerics}

\subsection{Numerical scheme to solve for \texorpdfstring{$\phi(x,t)$}{phi(x,t)}}

\begin{figure}[t]
\centering
\includegraphics[width=0.5\textwidth]{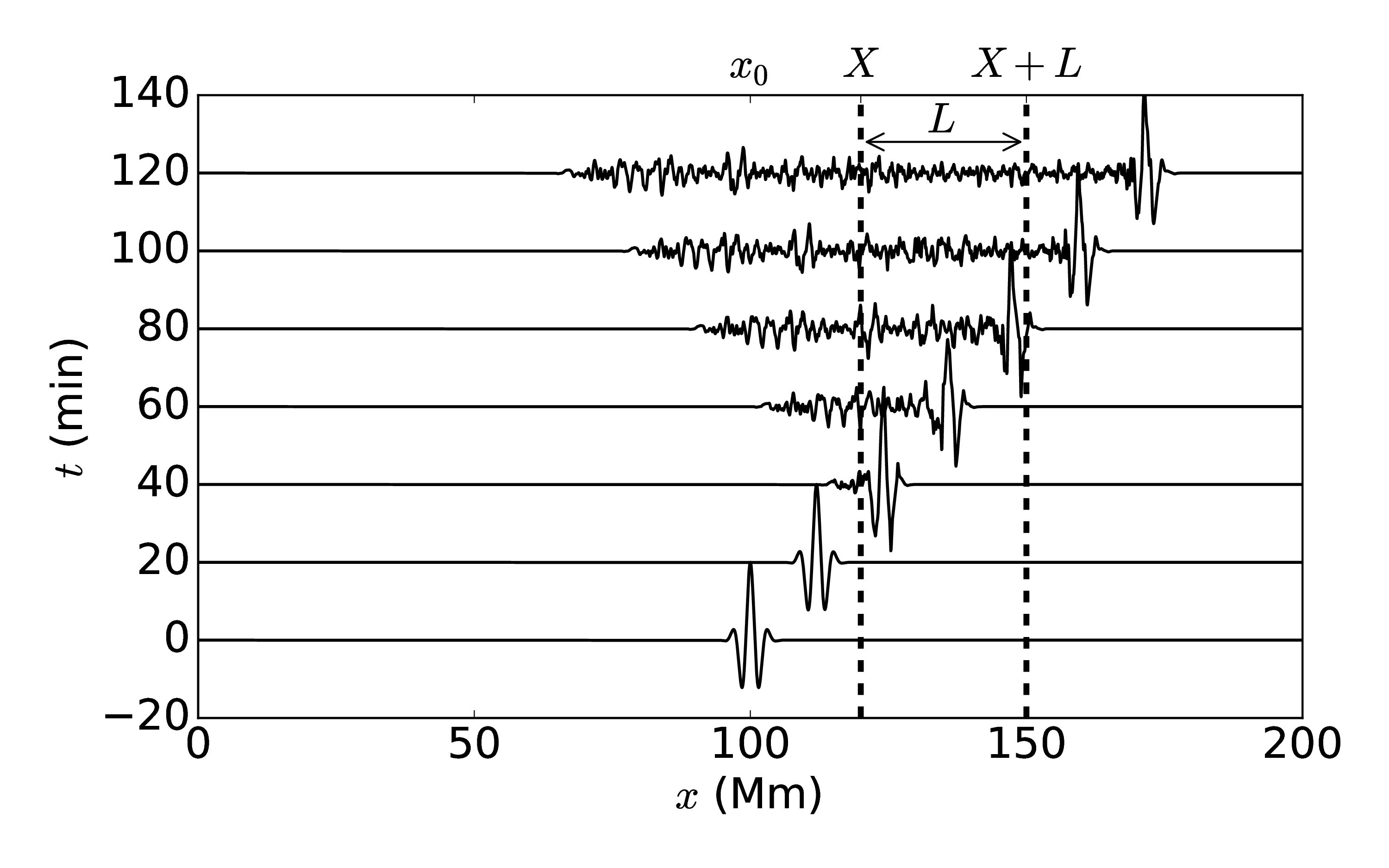}
\includegraphics[width=0.5\textwidth]{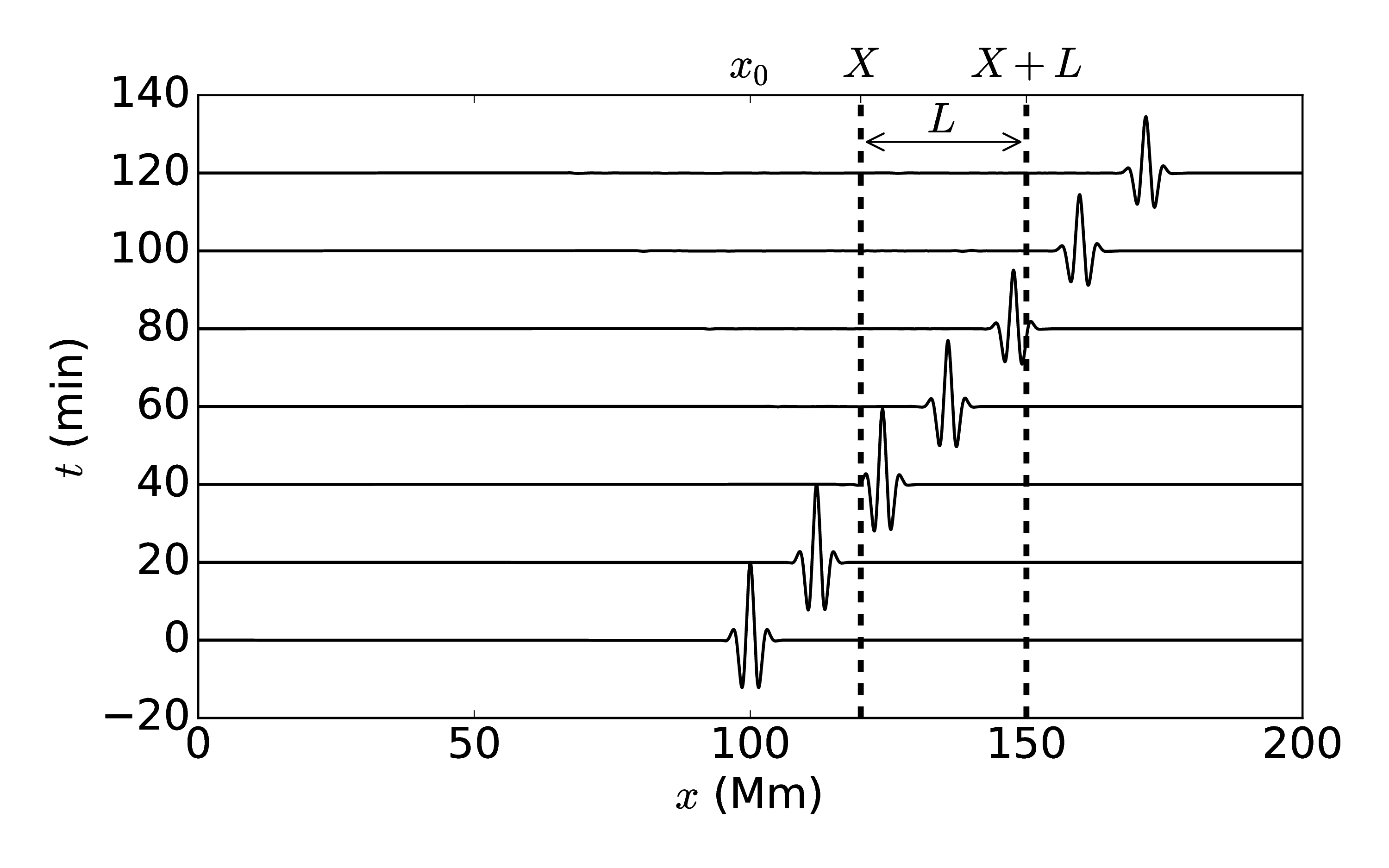}
\includegraphics[width=0.5\textwidth]{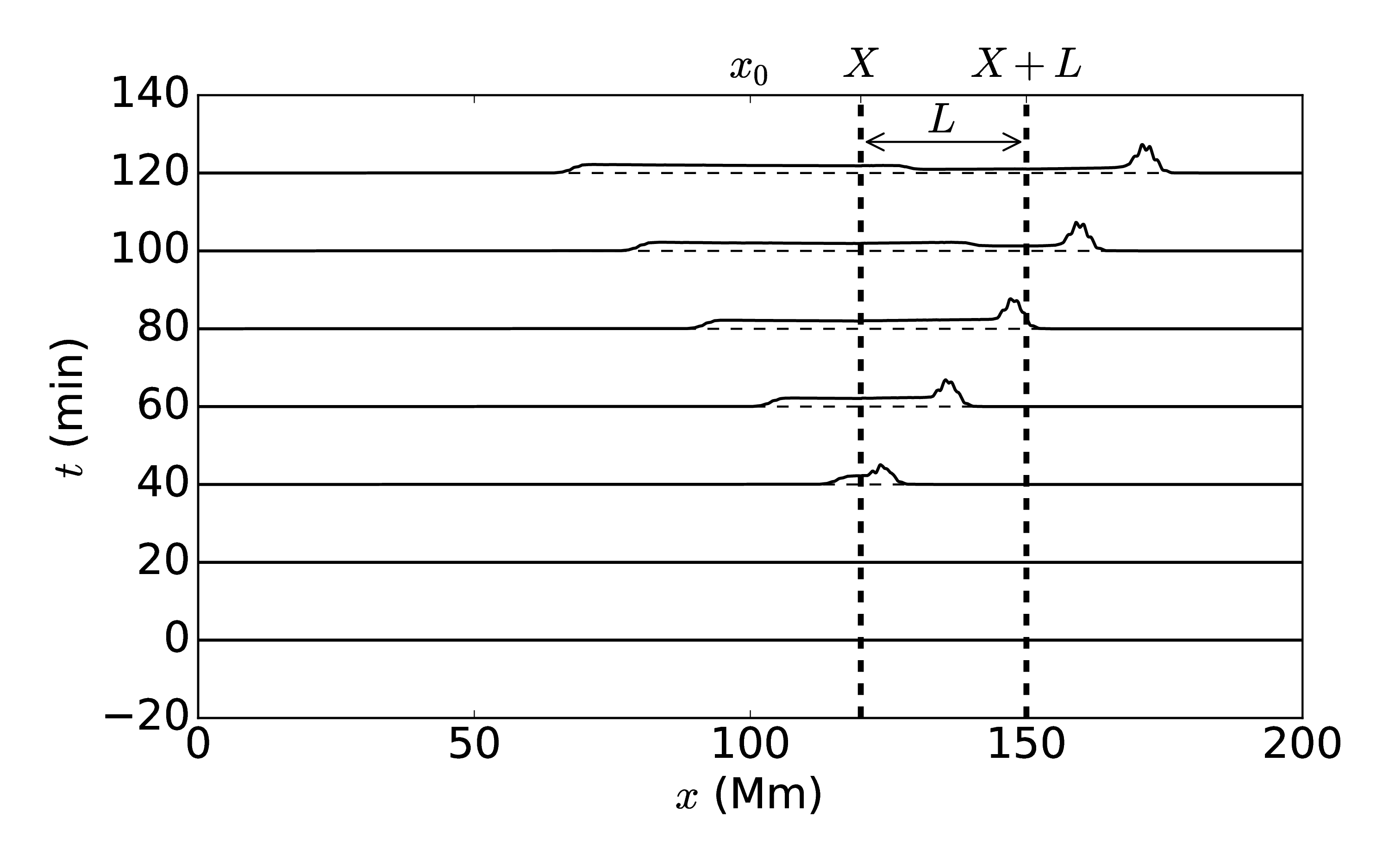}
\caption{Top: Wave packet propagation through a realization of a random medium (medium 2, with $\epsilon=0.1$ and $\tau=400$~s) located between the vertical dashed lines at different time steps. Middle: Average over $10\ 000$ realizations. Bottom: Square root of the variance of the wave field. See the movie online at \href{https://edmond.mpdl.mpg.de/imeji/collection/RXxuh3ldPBZ9bdNZ/item/qENjK0Bn_FscJUTr?q=&fq=&filter=&pos=0\#pageTitle}{Movie 1}.}
\label{propagation}
\end{figure}

In order to solve numerically Eq.~\eqref{bible_equation}, we use an explicit finite-difference scheme of second order. We choose $\omega_0/2\pi = 3$~mHz and $\sigma/2\pi = 1$~mHz so that the frequency range of study is $1$ to $5$~mHz, which is a reasonable choice for solar acoustic waves. The wave packet is initially at $x_0=100$~Mm, while $x_{\rm max}=200$~Mm and $t_{\rm max}=10000$~s. We set $X=120$~Mm and $L=30$~Mm. The resolutions for the simulations are $\Delta x=50$~km and $\Delta t=2.5$~s, so that $c_0 \Delta t/\Delta x=0.5 < 1$.

An example of time-domain simulation with medium 1 is shown in the online \href{https://edmond.mpdl.mpg.de/imeji/collection/RXxuh3ldPBZ9bdNZ/item/qENjK0Bn_FscJUTr?q=&fq=&filter=&pos=0\#pageTitle}{movie} and in Fig.~\ref{propagation}. The wave packet begins to be perturbed when it enters the random medium. Most of the signal is transmitted forward roughly in the form of a wave packet (ballistic wave packet). Small oscillations trail that signal, propagating either forward or backward. Once out of the perturbation, the shape of the wave packet is not modified anymore.

\subsection{Numerical scheme to solve for \texorpdfstring{$\tilde{\phi}(x,\omega)$}{phi(x,omega)}}

The code uses a second-order discretization scheme with a spatial resolution $\delta x = 4$~km. A tridiagonal system is inverted with the tridiagonal matrix algorithm (Thomas algorithm). $c_0$, $x_{\rm max}$, $X$ and $L$ are the same as for the time-domain code. The resolution is done for frequencies between $1$ and $5 \ {\rm mHz}$.

\subsection{Measuring the  attenuation} \label{sect:MeasAtt}

\begin{figure}[ht]
\includegraphics[width=0.5\textwidth]{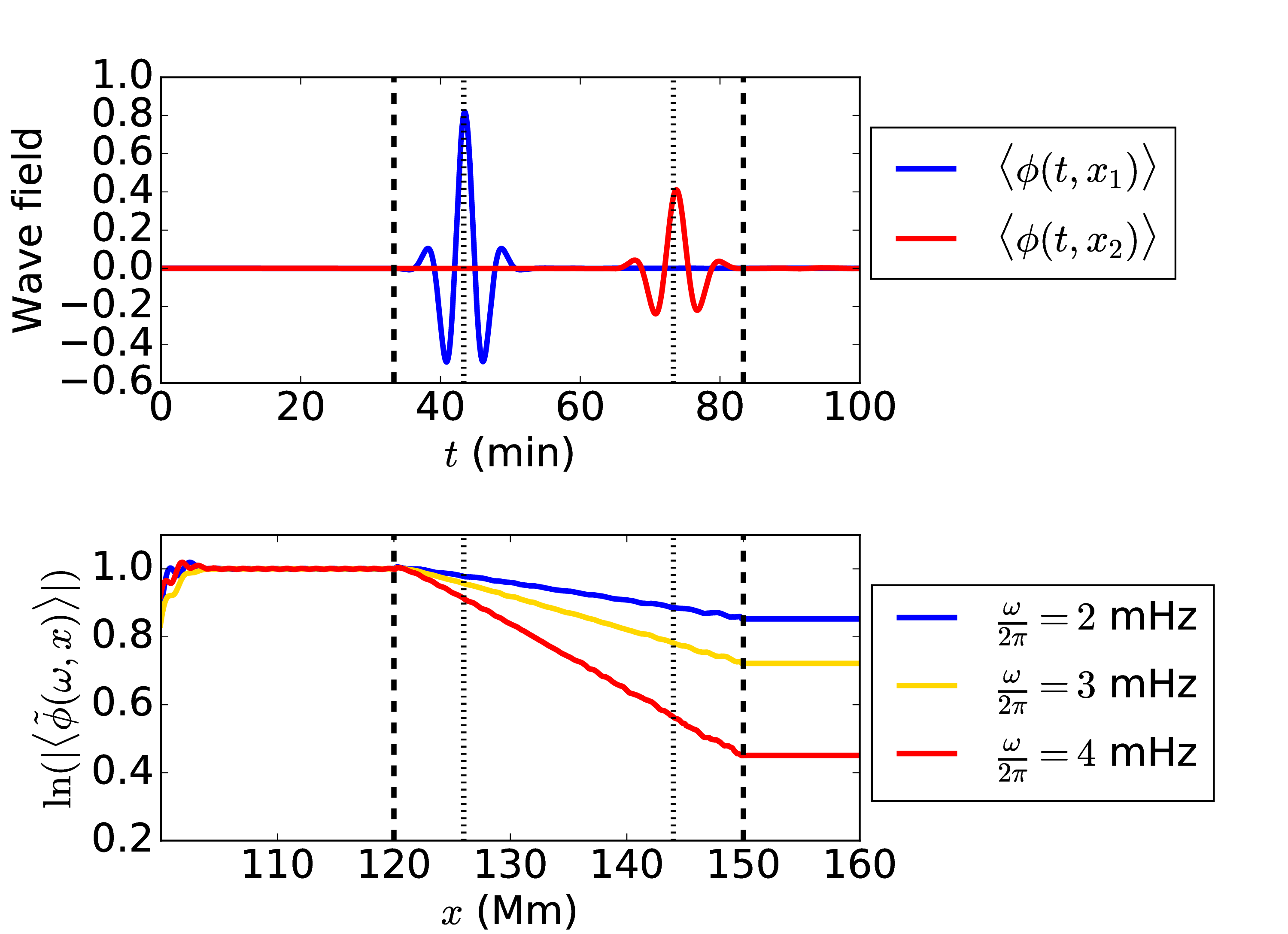}
\caption{Measuring the attenuation with the temporal code. Top: Coherent wave packet at $x_1=126$~Mm (blue) and $x_2=144$~Mm (red). Bottom: Natural log of the spectrum of the coherent wave packet at different frequencies. We fit its slope between the vertical dotted lines, corresponding to $[x_1,x_2]$. The vertical dashed lines delimit the location of the perturbation in time (for a wave packet propagating at $c_0$) and in space. In the figure, the Fourier components have been normalized so that they have the same amplitude before entering the perturbation.}
\label{measurement_method_attenuation_temporal_code}
\end{figure}

Following \citet{2002quse.book.....A}, after propagating between two points $x_1$ and $x_2$ ($x_2>x_1$) in an attenuating medium, a plane wave is damped by a factor 
$$e^{-k_i (x_2-x_1)}.$$
The spatial attenuation $k_i$ could be measured from the amplitude difference between the incoming and the outgoing wave packets. However, this method leads to artifacts due to boundary effects occurring at the edges of the random medium. Therefore, we rather consider the wave packet inside the perturbation. We take $x_1=126$~Mm and $x_2=144$~Mm, each point being $6$~Mm away from the edge of the perturbation. As shown in Fig.~\ref{measurement_method_attenuation_temporal_code}, we take the temporal Fourier transform of $\langle \phi(x,t) \rangle$ where $x\in[x_1,x_2]$. The power of the signal has been attenuated during the propagation from $x_1$ to $x_2$. At each frequency, we then fit a first order polynomial to the natural logarithm of the norm of the Fourier component in order to retrieve the decay coefficient.

\subsection{Measuring the effective wave speed} \label{sect:MeasC}

\begin{figure}[t]
\includegraphics[width=0.99\linewidth]{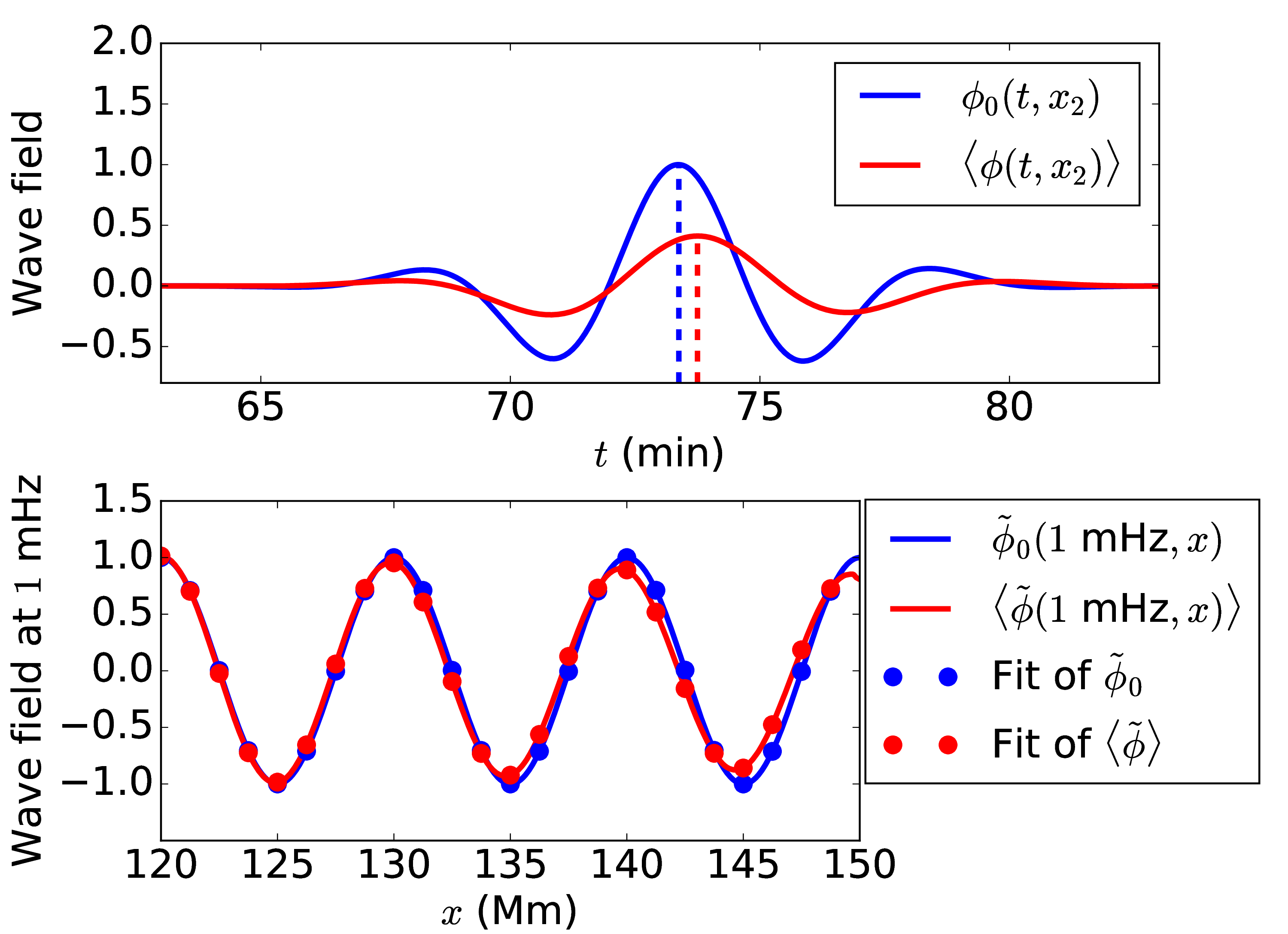}
\caption{Measuring the effective wave speed. Top: Coherent wave packet (red curve) experiencing a travel time shift compared to the unperturbed wave packet (blue curve). The vertical dashed lines represent the arrival times at $x_2$. Bottom: Fit of a decaying cosine to the real part of the temporal Fourier transform in the random region, at $\omega/2\pi=1$~mHz.}
\label{measurement_method_sound_speed_temporal_code}
\end{figure}

We first apply a temporal Fourier transform to $\langle \phi \rangle$. Then we fit to $\text{Re}(\langle \tilde \phi(x,\omega) \rangle)$ (we could have chosen the imaginary part arbitrarily), in the perturbed region, an exponentially decreasing oscillatory function where the decay rate has been determined via the method to measure the attenuation from the previous section. More precisely, we fit
\begin{equation}
A e^{-k_i(x-X)}\cos\left(\frac{\omega}{c_{\rm eff}(\omega)}(x-x_s)\right)
\end{equation}
where $A$, $x_s$, and $c_{\rm eff}$ are the free parameters, with $x_s$ being a phase shift. A similar fit is done on $\phi_0$ to take numerical dispersion into account. This is illustrated in Fig.~\ref{measurement_method_sound_speed_temporal_code}.

\section{Effective medium theories \label{sect:theories}}

\begin{table*}[t]
\begin{center}
\caption{Theories used in this paper for the effective wave speed and attenuation in a frozen medium as $\epsilon \rightarrow 0$, and their range of validity. The Keller and the Born theories are for medium 1. \label{tab:theory}}
\begin{tabular}{cccc}
\hline \hline
Theory & Validity range & $k_i$ & $c_{\rm eff}$ \\
\hline
Keller 1964 & $\epsilon \ll 1$ & $\epsilon^2 k_0 \left(k_0 a +\frac{k_0 a}{1+4(k_0 a)^2}\right)$ & $c_0\left(1-\frac{\epsilon^2}{2}\left[3-\frac{4(k_0a)^2}{1+4(k_0a)^2}\right]\right)$ \\
Born (2nd order) & $\epsilon^2(k_0 a)^2\frac{L}{a} \ll 1^{(\dag)}$ & $\simeq^{(\ddag)} \epsilon^2 k_0 \left(k_0 a +\frac{k_0 a}{1+4(k_0 a)^2}\right)$ & $\simeq^{(\ddag)} c_0\left(1-\frac{\epsilon^2}{2}\left[3-\frac{4(k_0a)^2}{1+4(k_0a)^2}\right]\right)$ \\
Homogenization & $k_0a \ll 1$ & Not applicable & $c_{\rm h}=\langle c^{-2}\rangle^{-1/2} \simeq c_0\left(1-\frac{3}{2}\epsilon^2\right)$ \\
Geometrical optics & $k_0a \gg 1$, $k_0a \gg 2\pi \frac{L}{a}$, $\epsilon \ll 1^{(\#)}$ & Not applicable & $c_{\rm ray}=\langle c^{-1} \rangle^{-1} \simeq c_0(1-\epsilon^2)$ \\
\hline
\end{tabular}
\end{center}
\tablefoot{$ ^{(\dag)}$ Approximation for $k_0a>1$ of \citet{1989psr3.book.....R} who made the derivation for a Gaussian autocorrelation function and single scattering. $ ^{(\ddag)}$ The dominant term is that of Keller for small perturbations (see for example Fig.~\ref{comparison_theories_simu}). $ ^{\#}$ See \citet{1989psr4.book.....R}.}
\end{table*}

Depending on the values of the parameters, in particular $k_0 a$ and $\epsilon$, different theories can be used to compute the effective parameters $c_{\textrm{eff}}$ and $k_i$. For a medium whose spatial scale is much less than the wavelength ($\lambda \gg a$, regime of Rayleigh scattering), the homogenization method is appropriate and gives an effective sound speed $c_{\rm eff} = c_0(1-3/2\epsilon^2)$ (derivation for a frozen medium in Appendix \ref{Spatial_homo_section}). On the other hand, for small wavelengths ($\lambda \ll a$, small-angle scattering regime), the geometrical optics approach is relevant and implies $c_{\rm eff} = c_{\rm ray} = c_0(1-\epsilon^2)$ (derivation for a frozen medium in Appendix \ref{Geom_optics_section}). These two approaches give an effective wave speed which is independent of frequency and of the power spectrum of the perturbation but do not provide any attenuation. Another caveat is that in our model, we are in the regime of Mie scattering or large-angle scattering \citep{1988sasw.book.....A} because $\lambda \simeq 3.3 a$. The wave number can therefore not be considered large nor small compared to $1$ and other theories may be required.

We explore two other derivations in the case of small perturbations ($\epsilon \ll 1$). In this regime, two methods are used: the Keller solution, derived for a frozen (Appendix \ref{Keller_section}) or time-dependent medium (Appendix \ref{Keller_temporal_section}); the Born second-order approximation for a frozen medium (derived in Appendix \ref{Born_section}). The Keller and the Born solutions converge toward the same values for small perturbations ($\epsilon=0.01$ for instance). However, at $\epsilon=0.1$, the Born solution is very different from the Keller one, and it is not possible to fit a function of the form of Eq.\eqref{representation}. We come back to this point in Section~\ref{sect:agreement}.

If $k_0 a \ll 1$ or $k_0 a \gg 1$, the effective wave speed obtained from the Keller theory converges toward the results of the homogenization and geometrical optics approaches, respectively. Table~\ref{tab:theory} summarizes the expressions for a frozen medium 1 (for which the analytical expressions can be easily derived) as $\epsilon \rightarrow 0$. We check the agreement of these theories with our simulations in Appendix~\ref{Comparison_theories_section}. The Keller and the Born second-order theories are consistent and predict that $k_i$ is proportional to $k_0^2$ while the effective wave speed difference is essentially independent of $k_0$ provided that $k_0 a \gtrapprox 1$. For comparison purposes, we note that \citet{1963ASR.12..223M} and \citet{2012swpshe.book.....S} (pp. 214-220) found that for a 3D frozen medium with an exponential autocorrelation (i.e., similar to medium 1), the attenuation is proportional to $k_0^4$ for $k_0 a \ll 1$ and to $k_0^2$ for $k_0 a \gg 1$ while the behavior of the effective wave speed is in qualitative agreement. In particular, the effective wave speed is always less than $c_0$. On the other hand, \citet{2001GeoJI.145..631V} used the wave localization theory to make use of so-called self-averaging quantities; he derived the effective medium using one realization of a one-dimensional perturbation in density and bulk modulus. He found that the attenuation coefficient tends toward a constant value at high $k_0 a$, while the effective wave speed difference is positive, in agreement with previous studies \citep[e.g.,][]{1992GeoJI.110...29M}. The sign of the difference $c_{\rm eff}-c_0$ and the dependence of the attenuation on frequency therefore seem to depend strongly on the equation that is solved.

\section{Results and comparisons \label{Results_section}}

 We compute the properties of the effective medium using the procedure explained in Sections~\ref{sect:MeasAtt}~and~\ref{sect:MeasC}, with $\epsilon=0.1$ and $a=1$~Mm. The $1$-$\sigma$ error bars shown later on on the attenuation and wave speed measurements are obtained from ten sets of $10^4$ realizations.

\subsection{Coherent wave field \label{sect:agreement}}

\begin{figure}[ht]
\includegraphics[width=0.5\textwidth]{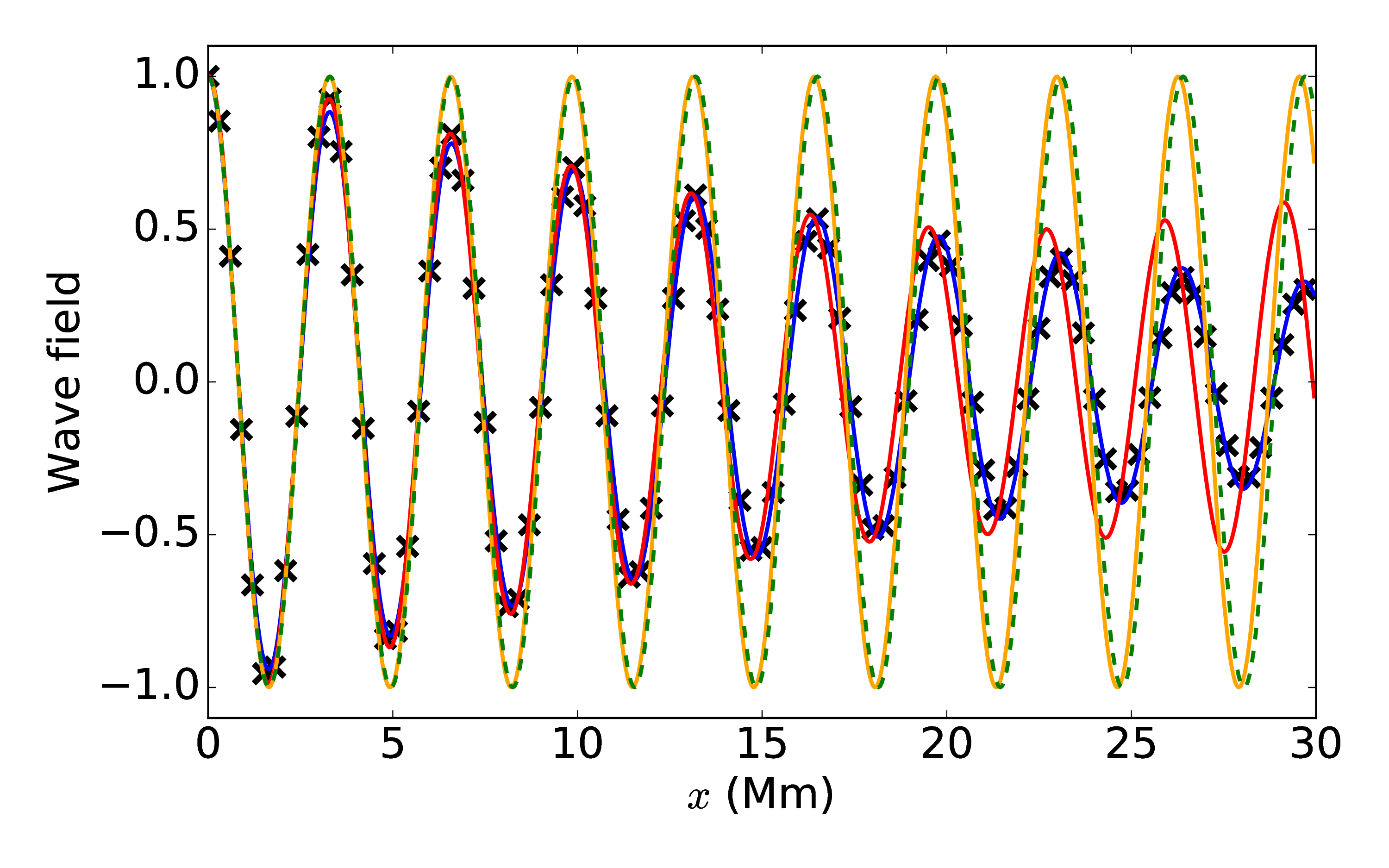}
\caption{Theoretical solutions compared with the coherent wave field inside the random medium from the numerical simulation, for a frozen medium 1, at $3 \ {\rm mHz}$. Keller: blue. Born: red. Homogenization: orange. Geometrical optics: dashed green. The black crosses are the numerical solution.}
\label{plot_all_theories}
\end{figure}

We first reconstruct the theoretical coherent wave field obtained when using the various theories. For all of them (except the Born theory that provides directly the wave field), we assume the form written in Eq.~\eqref{representation}. In Fig.~\ref{plot_all_theories}, we plot $\text{Re}(\langle \tilde \phi(3\ {\rm mHz},x)\rangle)$ ($x \in [X,X+L]$) for a frozen medium 1. Clearly the Keller approximation does the best job at approximating the true (numerical) solution. As mentioned before, neither the homogenization technique nor the geometrical optics makes an attenuation emerge. The Born solution is a good approximation on about half of the random medium at this frequency, before it becomes out of phase with the numerical solution while its amplitude also starts to increase. The discrepancy is worse and arises earlier in the medium for higher frequencies. Thus the Born approximation, although similar to the Keller approximation when $\epsilon \rightarrow 0$, performs poorly for a $10\%$ perturbation in a medium of size $L=30 \ {\rm Mm}$.

\subsection{Attenuation}

\begin{figure}[ht]
\includegraphics[width=0.5\textwidth]{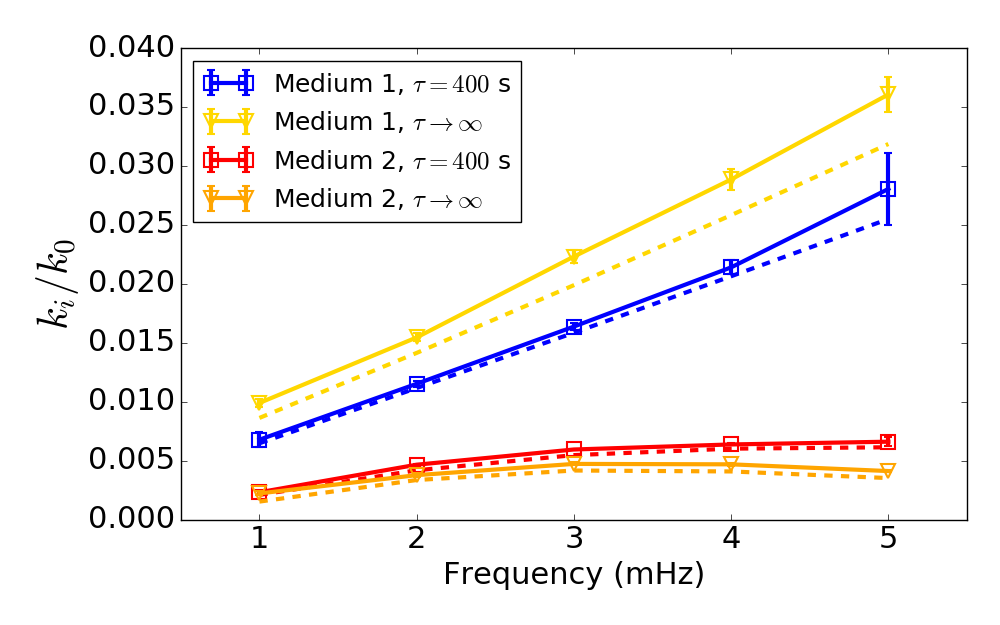}
\caption{Attenuation of the coherent wave packet vs frequency for media 1 and 2, after propagation through a band of perturbed medium. The 1D theory from Keller is overplotted in dashed lines. 1-$\sigma$ error bars are shown.}
\label{attenuation_vs_freq}
\end{figure}

Fig.~\ref{attenuation_vs_freq} shows the measured attenuation for simulations with $\tau=400$~s and $\tau \rightarrow \infty$. The case $\tau=1$~day (not shown on the plot) lies within the error bars of the curve for the frequency code, which is to be expected as the typical time scales involved (the period of the wave, about $5$~minutes, and the time it takes for it to travel through the medium, about $1$~h) are much less than one day. We superimpose the attenuation that one expects from the time-dependent Keller theory.

The attenuation by medium 1 is an increasing function of frequency, with a value of about $1.5\%$ of the wave number at $3 \ {\rm mHz}$ for $\tau=400 \ {\rm s}$. The ratio $k_i/k_0$ is a linear function of frequency, meaning that $k_i$ is quadratic, as expected from the Keller theory. For medium 2 however, the attenuation is not quadratic. It reaches about $0.5\%$ of the wave number at $3 \ {\rm mHz}$, which is smaller than the medium 1 value by a factor $3$. The smaller attenuation values are caused by the lack of power toward low wave numbers in the spectrum of the perturbation: the absence of large scales in the perturbation means that the incoherence between the realizations of the wave packets occurs preferentially on small scales, thereby decreasing the overall broadening of the wave packet. In medium 2, the ratio $k_i/k_0$ stabilizes above $5$~mHz for $\tau=400$~s, while it reaches a maximum at about $3$~mHz for $\tau\rightarrow\infty$. This may indicate that there is a preferred scale of damping of the coherent wave field.

\subsection{Effective wave speed}

\begin{figure}[t]
\includegraphics[width=0.5\textwidth]{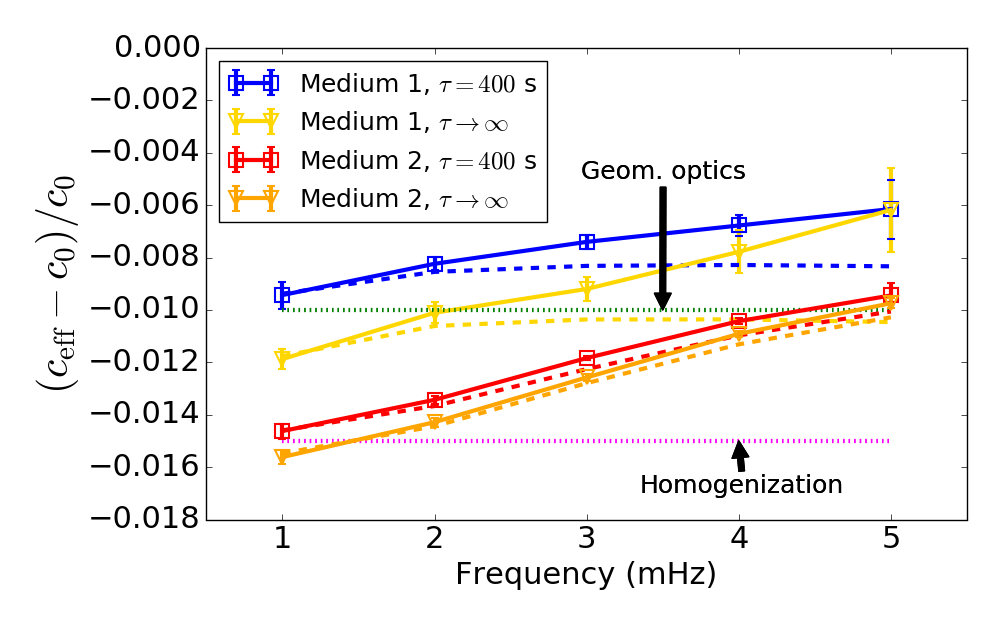}
\caption{Speed of the coherent wave packet vs frequency for media 1 and 2, after propagation through a perturbed medium. The 1D theory adapted from Keller (dashed lines) as well as the prediction from the homogenization theory and the geometrical optics are shown. 1-$\sigma$ error bars are shown.}
\label{c_vs_frequency_eps01}
\end{figure}

Fig.~\ref{c_vs_frequency_eps01} shows the effective wave speed computed with for $\tau=400$~s and $\tau \rightarrow \infty$, as well as the time-dependent Keller theory, the (frozen) spatial homogenization solution and the (frozen) geometrical optics solution. Like for the attenuation, the case $\tau=1$~day (not shown) lies within the error bars of the curve for the frequency code. The effective wave speed is less than the unperturbed sound speed $c_0$. This is due in part to waves being scattered back and forth, contributing to the overall transmitted signal but at a later time than the unperturbed wave. The second reason is the delay experienced by forward-scattered waves. Indeed, in the regime of geometrical optics ($\lambda/a\ll 1$) where scattering occurs essentially forward, the effective wave speed is given by the geometric velocity $c_{\rm ray}=\langle c^{-1} \rangle ^{-1}<c_0$. 

The effective wave speed in medium 1 is an increasing function of frequency, with a shift from $c_0$ by about $-0.7\%$ at $3 \ {\rm mHz}$ for $\tau=400 \ {\rm s}$. The Keller theory is in relative agreement for low frequencies ($f \leq 2 \ {\rm mHz}$) but it predicts a constant wave speed at higher frequencies. On the other hand, the measured effective wave speed in medium 2 clearly changes from the homogenized velocity $c_{\rm h}$ at $1$~mHz to the geometric velocity $c_{\rm ray}$ at $5$~mHz. We note a remarkable agreement at all frequencies between the simulations and the Keller theory for medium 2.

\subsection{Variance of wave field \label{sect:var}}

\begin{figure}[ht]
\includegraphics[width=0.5\textwidth]{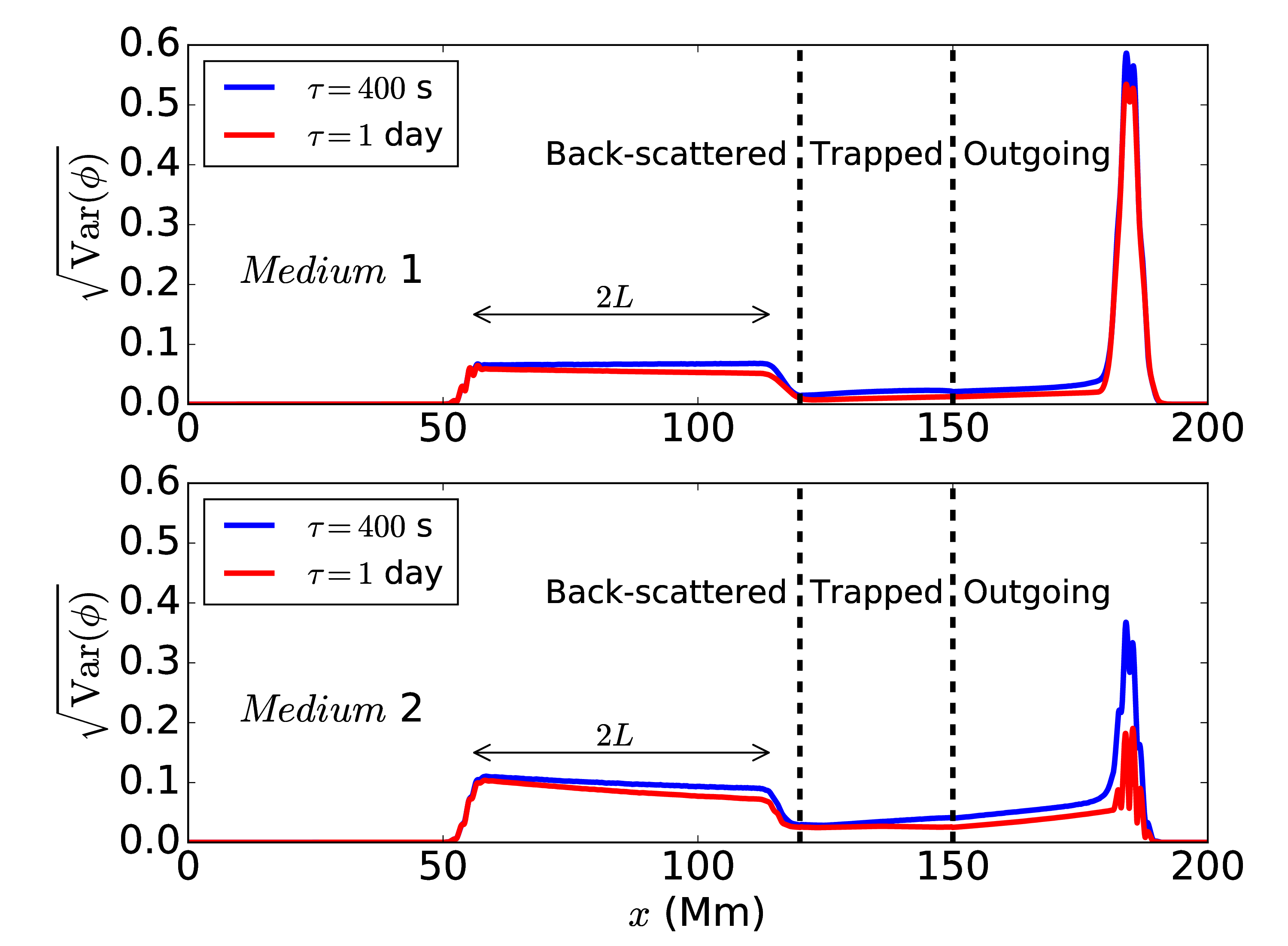}
\caption{Square root of the variance of the wave field as a function of position at a given time $t=8500$~s. Top: Medium 1. Bottom: Medium 2. See the movie online at \href{https://edmond.mpdl.mpg.de/imeji/collection/RXxuh3ldPBZ9bdNZ/item/t5_XVSoqh2E0iWmo?q=&fq=&filter=&pos=1\#pageTitle}{Movie 2}.).
}
\label{dispersion_vs_x_withgranulation}
\end{figure}
The mean of the perturbation is zero, therefore looking at the coherent wave field may not be enough to directly detect multiple scattering because one would only see oscillations mixed within the noise. In the regime of strong perturbations, the coherent part would vanish and only the fluctuating part would remain, solely accessible via second order moments. One can for instance look at the envelope of the signal by studying the variance of the wave field. 

As shown in Fig.~\ref{dispersion_vs_x_withgranulation} and in the online \href{https://edmond.mpdl.mpg.de/imeji/collection/RXxuh3ldPBZ9bdNZ/item/t5_XVSoqh2E0iWmo?q=&fq=&filter=&pos=1\#pageTitle}{movie}, it is composed of three parts: a peak corresponding to the variance of the ballistic wave packet, coda waves (late-arriving waves) propagating forward, and coda waves propagating backward. The forward-propagating coda results from waves back-scattered an even number of times in the perturbed medium. The backward-propagating coda forms a plateau of width $2L$ and results from single back-scattering. In geophysics, a connection has been made between the functional form of the coda in time domain and the complexity of the scattering medium \citep[e.g.,][]{2012swpshe.book.....S}.

We decompose the domain in three regions (before, after and in the random medium) and integrate spatially the variance over each of these three regions at $t_m=8500$~s, i.e. after the coherent wave packet went through the random medium and just after the plateau of back-scattered signal went out of it:
\begin{align}
    &\mathcal{E}_{bsc} = \int_0^X Var(\phi(x,t_m)) \ \mathrm{d}x,  \\
    &\mathcal{E}_{out} = \int_{X+L}^{x_{\rm max}} Var(\phi(x,t_m)) \ \mathrm{d}x, \\
    &\mathcal{E}_{tr} = \int_X^{X+L} Var(\phi(x,t_m)) \ \mathrm{d}x.
\end{align}
It gives us a measurement of the variance that, respectively, has been back-scattered, transmitted or is still trapped in the slab at this particular time. For medium 2, the back-scattered variance makes up for about $50\%$ of the total variance for $\tau=400$~s, and $75\%$ for $\tau=1$~day. The reason for these high amounts is that the spectrum of medium 2 peaks at small scales, therefore more back-scattering takes place than for instance in medium 1 where these values become respectively $20\%$ and $15\%$.

\subsection{Dependence on correlation time of the medium} \label{sect:departure}

Calculations of an effective medium are easier to carry when the perturbation is frozen because one can work directly in the frequency domain. Therefore, we study here how the effective parameters $k_i$ and $c_{\rm eff}$ depend on the correlation time of the medium.

\begin{figure}[ht]
\includegraphics[width=0.5\textwidth]{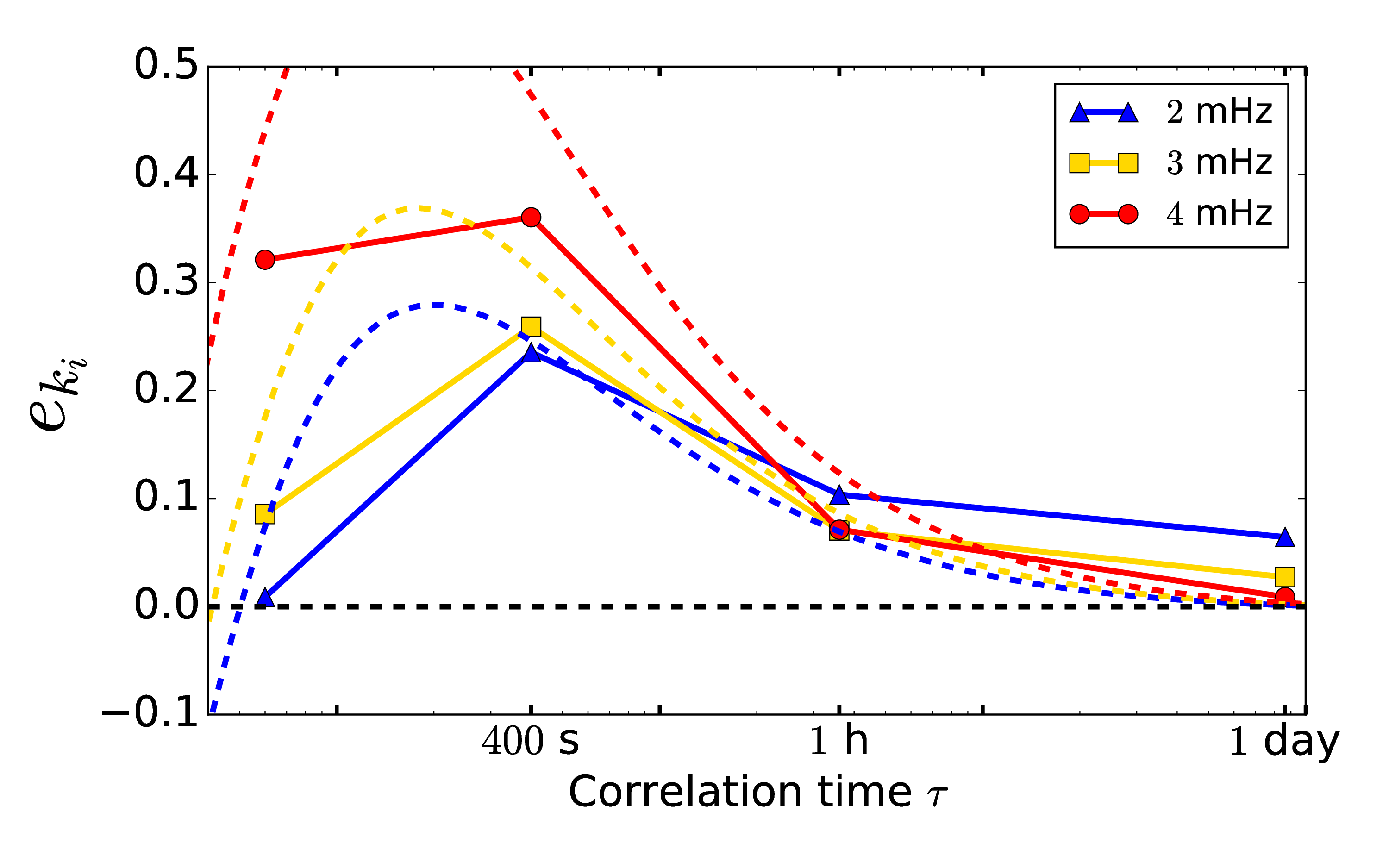}
\includegraphics[width=0.5\textwidth]{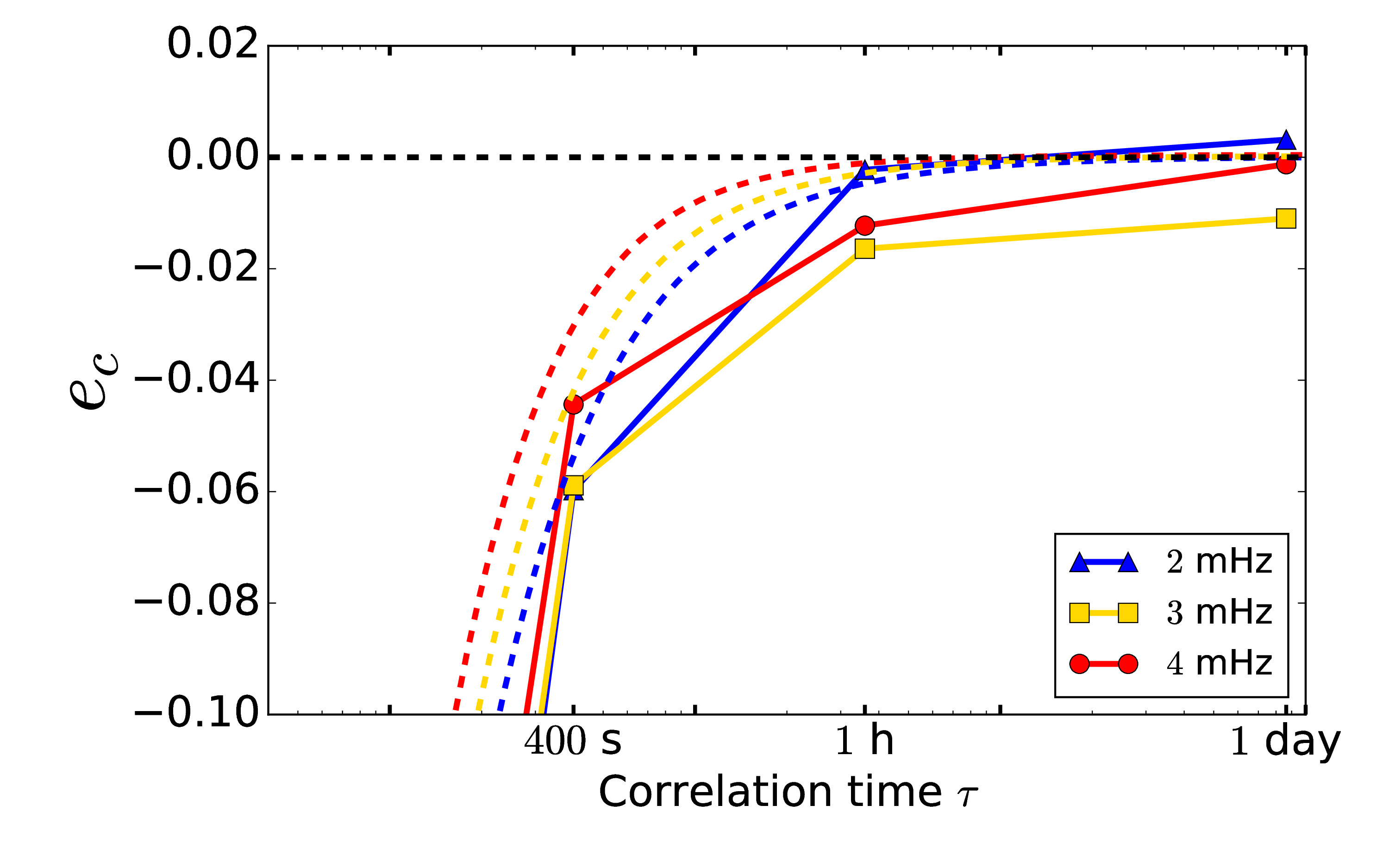}
\caption{Relative error on the attenuation (top) and the effective wave speed difference (bottom) at $2$, $3$ and $4$~mHz (medium 2).  The error is between the quantities at $\tau$ and at $\tau\rightarrow \infty$. The dashed lines are the predictions from the time-dependent Keller theory.}
\label{relative_error_average_withgranulation}
\end{figure}

Fig.~\ref{relative_error_average_withgranulation} shows the relative errors in the attenuation, $e_{k_i}$, and in the effective wave speed difference, $e_c$, between a given correlation time and the $\tau \rightarrow \infty$ case at $2$, $3$ and $4$~mHz, for medium 2:
\begin{align}
    e_{k_i}(\omega,\tau) &= \frac{k_i(\omega,\tau)-k_i(\omega,\infty)}{k_i(\omega,\infty)}, \\
    e_c(\omega,\tau) &= \left( \frac{c_{\rm eff}(\omega,\tau)-c_0}{c_0} - \frac{c_{\rm eff}(\omega,\infty)-c_0}{c_0} \right)\left( \frac{c_{\rm eff}(\omega,\infty)-c_0}{c_0}
    \right)^{-1} \nonumber \\
    &=\frac{c_{\rm eff}(\omega,\tau)-c_{\rm eff}(\omega,\infty)}{c_{\rm eff}(\omega,\infty)-c_0}.
\end{align}
$e_{k_i}$ being generally positive, the attenuation is underestimated by the frozen-medium approximation. Our understanding is that since the power of the perturbation mostly lies at high wave numbers, the attenuation mostly comes from the small-scale incoherence between the realizations of the wave packets. Therefore, there must be two regimes: one at small values of $\tau$ where the attenuation increases with $\tau$, and one at greater values of $\tau$ where the attenuation decreases, because persisting scatterers start to create less small-scale incoherence, so less attenuation. The transition between the two regimes corresponds to a resonance, located according to the theory at about $\tau=195$~s, $\tau=180$~s and $\tau=135$~s at $2$, $3$ and $4$~mHz. On the other hand, $e_c$ being negative, the approximation overestimates the decrease in effective wave speed, because longer-lived features are better "seen" by the wave packets. The decrease is therefore a monotonic function of $\tau$, with its asymptotic value at $\tau \rightarrow \infty$ only determined by the value of the ratio of the wave number over the typical size of the scatterer. The error is frequency-dependent and, on average over the three central frequencies, is $29\%$ (respectively $-5\%$) for the attenuation (respectively the effective wave speed difference) at $\tau=400$~s.

\begin{figure}[ht]
\includegraphics[width=0.5\textwidth]{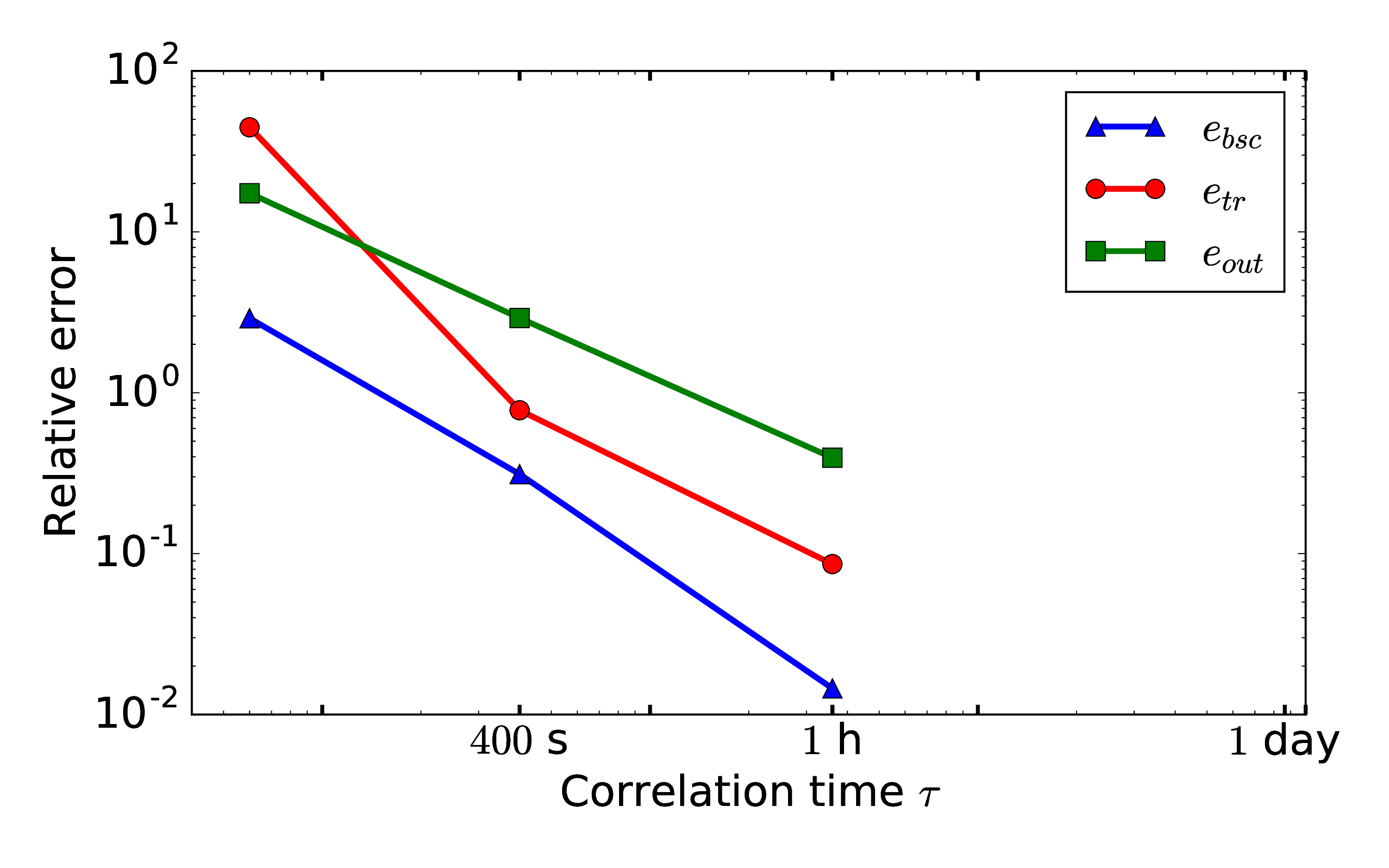}
\caption{Relative error in the variance integrated in space before, in and after the random medium, at $t=8500$~s. The error is between the quantities at $\tau$ and at $\tau=1$~day.
}
\label{relative_error_dispersion_granulation}
\end{figure}
As for the variance, we assume $\tau = 1 \; \text{day} \simeq \infty$. This is justified as the propagation time in the random medium of length $30$~Mm is about $1 \;\text{h} \ll 1 \;\text{day}$. We compute therefore
\begin{align}
    &e_{bsc}(\tau) = (\mathcal{E}_{bsc}(\tau)-\mathcal{E}_{bsc}(1 \ {\rm day}))/\mathcal{E}_{bsc}(1 \ {\rm day}), \\
    &e_{out}(\tau) = (\mathcal{E}_{out}(\tau)-\mathcal{E}_{out}(1 \ {\rm day}))/\mathcal{E}_{out}(1 \ {\rm day}), \\
    &e_{tr}(\tau) = (\mathcal{E}_{tr}(\tau)-\mathcal{E}_{tr}(1 \ {\rm day}))/\mathcal{E}_{tr}(1 \ {\rm day}).
\end{align}
The relative errors at $\tau=400 \ {\rm s}$ are then about $30\%$, $80\%$ and $290\%$ for the back-scattered, trapped and outgoing variance, respectively. Hence it appears that for medium 2, the variance is more sensitive to the correlation time than the coherent wave field, and that the back-scattered coda is less sensitive than the rest of the variance.

The relative errors for medium 1 at $\tau=400 \ {\rm s}$ are presented for comparison purposes in Table~\ref{tab:results}. In this case, the frozen-medium approximation overestimates the attenuation. Most of the power is indeed located at large scales, so the attenuation is mostly caused by the large-scale incoherence between the realizations (shifts of the wave packets), which triggers a broadening and damping of the coherent wave packet. Therefore, the impact of scattering is larger if the scatterers persist while the wave packets propagate through them than if the scatterers evolve in time. On the other hand, the frozen-medium approximation still overestimates, albeit by a larger amount, the decrease in effective wave speed. The error for both quantities does not depend much on frequency, and is about $-25 \%$ for the attenuation and $-19 \%$ for the effective wave speed. For the back-scattered coda, the error increases to $46\%$.

\begin{table}
\centering
\caption{Relative error at $\tau=400 \ {\rm s}$ for the measured quantities for media 1 and 2. $e_{k_i}$ and $e_c$ are averaged over the three central frequencies.
}
\begin{tabular}{cccc}
\hline \hline \\
   & & Medium 1 & Medium 2 \\
     \hline
   \multirow{2}{*}{Coherent wave} & $e_{k_i}$ & $-25\%$ & $29\%$\\
    & $e_c$  & $-19\%$ & $-5\%$\\
    \hline
    Variance & $e_{bsc}$ & $46\%$ & $31\%$ \\
\end{tabular}
\tablefoot{For the coherent wave field, the errors are computed using both the temporal and the frequency codes. They are averaged over the three central frequencies $2$, $3$ and $4$~mHz. On the other hand, since we study the variance in time domain, the errors for this quantity are computed using only the temporal code.}
\label{tab:results}
\end{table}

\section{Discussion \label{sect:discussion}}

\subsection{Accuracy of the theories}

All theories predict a decrease in the effective wave speed. The effective wave speed and the attenuation of the coherent wave field are best described by the Keller approximation. The Born second-order solution, although consistent with the Keller solution for small perturbations, performs poorly for larger amplitudes, therefore it may not be suited for the study of acoustic wave scattering by solar granulation unless it is on small distances ($<30 \ {\rm Mm}$). The homogenization technique and the geometrical optics do not model the attenuation of the coherent wave field. However they correctly represent the decrease in wave speed for low and high frequencies, respectively.

\subsection{Validity of the frozen-medium approximation}

It is more convenient to study acoustic wave propagation in the frequency domain, but this is easily doable only when the coefficients of the wave equation do not depend on time, i.e. when one can use a snapshot of the random medium. As summarized in Table~\ref{tab:results}, we find that for medium 2, the attenuation is underestimated by the frozen-medium approximation by $29\%$ at the frequencies of interest for the Sun. As for the effective wave speed difference, which is an important quantity since it is directly related to the helioseismic travel times, it is overestimated by $5\%$. The greater error for $k_i$ seemingly arises from the presence of a resonance of the function $k_i(\tau)$ at a correlation time close to that of granulation, while the effective wave speed does not exhibit such a feature. We note that the relative error in $c_{\rm eff}-c_0$ is similar to that of $k_i$ in medium 1, when the power of the perturbation is distributed at low scales. The frozen-medium approximation underestimates the variance of the amplitude of back-scattered coda waves by about $30\%$.

\subsection{Detectability of coda waves}

The numerical simulations show the emergence of coda waves, which are an interesting effect of multiple scattering present both in single realizations of the wave field and in its variance, but not in the coherent wave field. Coda waves are seen trailing the ballistic wave packet, and also as late arrival  back-scattered waves (in one dimension). In helioseismology, acoustic waves are measured via the two-point cross-covariance function of the solar oscillations. Therefore, in order to identify coda waves in the Sun, one needs to study the statistical variance of this cross-covariance function.

\begin{acknowledgements}
 We thank Aaron C. Birch for useful discussions and comments. PLP is a member of the International Max Planck Research School (IMPRS) for Solar System Science at the University of G\"ottingen. The computational resources were provided by the German Data Center for SDO through grant 50OL1701 from the German Aerospace Center (DLR). 
\end{acknowledgements}

\bibliographystyle{aa}
\bibliography{aanda}

\begin{appendix}

\section{Keller approximation: Time-independent random medium \label{Keller_section}}

Starting from a time-independent random medium $c(x)$,  we can take the Fourier transport of the wave equation:
\begin{equation}
\omega^2 \tilde \phi(x,\omega)+\partial_x^2(c^2(x) \tilde \phi(x,\omega)) = 0. \label{eq:waveFreq}
\end{equation}
The autocorrelation written in Eq.~\eqref{eq:autocor} can be simplified to
\begin{equation}
\langle \delta c(x) \delta c(x'+x) \rangle=c_0^2\epsilon^2f(x).
\end{equation}
For clarity, we drop the argument $\omega$ in the expression of $\tilde \phi$. 

\citet{1964PSAM.16..145M} considers an unbounded spatially random medium and assumes statistical homogeneity, isotropy and stationarity. The calculation could be generalized to the case of a localized perturbation, however we follow the original derivation. It does accurately model our problem since the amplitude attenuation and the effective wave speed shift arise because of the perturbed region. Therefore only the boundary effects are not taken into account.
Keller made the first part of his derivation in time-domain, using the fact that the Green's function for the 3D wave equation is essentially a delta function, which simplifies the calculation. In 1D however, the Green's function is related to the Heaviside step function. We shall first derive the Keller solution in frequency domain for a frozen medium, then generalize in Appendix \ref{Keller_temporal_section} to the solution in time domain.

The wave equation given by Eq.~\eqref{eq:waveFreq} can be written as
\begin{equation}
(\tilde L_0+\tilde L_1+\tilde L_2)\tilde \phi = 0,
\end{equation}
where  
\begin{eqnarray}
\tilde L_0 \tilde \phi &=& \omega^2 \tilde \phi +c_0^2\partial_x^2 \tilde \phi, \\
\tilde L_1 \tilde \phi &=& 2c_0 \partial_x^2 \Bigl( \delta c(x)\, \tilde \phi \Bigr), \\
\tilde L_2 \tilde \phi &=& \partial_x^2 \Bigl( \delta c^2(x)\, \tilde \phi \Bigr).
\end{eqnarray}
The unperturbed equation, assuming a constant background sound speed, is
\begin{equation}
\tilde L_0  \tilde \phi_0 = 0.
\end{equation}
The corresponding Green's function $G_0$, solution of $\tilde L_0 \tilde G_0(x,x') = \delta(x-x')$ where $\delta$ is the Dirac delta function, is
\begin{equation}
\tilde G_0(x,x') = -\frac{i}{2c_0^2k_0}e^{ik_0|x-x'|},
\end{equation}
where $k_0 = {\omega}/{c_0}$.
Keller has shown that one can find a new wave equation for the coherent wave field under the form
\begin{equation}
(\tilde L_0-\langle \tilde L_1 \tilde L_0^{-1} \tilde L_1\rangle+\langle \tilde L_2 \rangle) \langle \tilde \phi \rangle = 0, \label{eq:Bourret}
\end{equation}
with
\begin{align}
&(\langle \tilde L_1 \tilde L_0^{-1} \tilde L_1\rangle \langle \tilde \phi \rangle) (x) \nonumber \\
&= \left\langle c_0 \partial_x^2 \left( 2\delta c(x) \int_{-\infty}^\infty \mathrm{d}x' \tilde G_0(x,x')c_0\partial_{x'}^2[2\delta c(x') \langle \tilde \phi(x') \rangle ] \right)\right\rangle \nonumber \\
&= 4c_0^4\epsilon^2 \partial_x^2 \left( \int_{-\infty}^\infty \mathrm{d}x' \tilde G_0(x,x')\partial_{x'}^2[f(x'-x) \langle \tilde \phi(x') \rangle ] \right).
\label{second_order_term}
\end{align}
We assume that the coherent wave field also satisfies a wave equation with a complex wave number $k$ so that
\begin{equation}
\langle \tilde \phi(x') \rangle = e^{ikx'}.
\end{equation}
In this case,
\begin{equation}
\partial_{x'}^2[f(x'-x) \langle \tilde \phi(x') \rangle ] =
[ ( \partial_{x'}+ik)^2 f(x'-x)  ]
 e^{ikx'}   .
\end{equation}
Therefore,
\begin{align}
(\langle \tilde L_1 \tilde L_0^{-1} \tilde L_1\rangle \langle \phi  \rangle) (x) &= 4c_0^4\epsilon^2\partial_x^2 \left( e^{ikx} I(x)  \right) \nonumber\\
&= 4c_0^4\epsilon^2 \left(  (\partial_x+ik)^2 I(x) \right)  \,  \langle \phi(x) \rangle, \label{eq:Keller_term1}
\end{align}
where
\begin{equation}
I(x) = \int_{-\infty}^\infty \mathrm{d}x' \,  \tilde G_0(x,x')[(\partial_{x'}+ik)^2 f(x'-x)] e^{ik(x'-x)}.
\label{term_F}
\end{equation}
On the other hand,
\begin{equation}
\langle \tilde L_2 \rangle \langle \tilde \phi \rangle = -c_0^2 \epsilon^2 k^2 \langle \tilde \phi \rangle.
\label{not_to_forget}
\end{equation}
Using Eqs.~\eqref{eq:Keller_term1}~and~\eqref{not_to_forget} in Eq.~\eqref{eq:Bourret}, the perturbed wave equation for the coherent wave field is
\begin{equation}
\left( \partial_x^2 + k_0^2 - 4c_0^2\epsilon^2(\partial_x+ik)^2I(x) -  \epsilon^2 k^2 \right) \langle \tilde \phi(x) \rangle = 0.
\end{equation}
We can define the complex wave number by
\begin{equation}
k^2 = k_0^2 - 4c_0^2\epsilon^2(\partial_x+ik)^2I(x) - \epsilon^2 k^2.
\end{equation}
Since the autocorrelation function of the perturbation depends here only on the difference $x'-x$, $I(x)=I$. In the small-perturbation approximation, one can also replace $k$ by $k_0$ in the right-hand term, to get finally
\begin{equation}
k^2 = k_0^2(1+4c_0^2\epsilon^2I-\epsilon^2) .
\end{equation}
We note that it is possible to keep $k$ in the right-hand side, one then has to solve a biquadratic complex equation. Here we only use the approximation. 

In this paper, we used in one case an exponential correlation function
\begin{equation}
f_1(x'-x)=f_1(\zeta) = \epsilon^2e^{-|\zeta|/a},
\end{equation}
where $\zeta = x'-x$. In this case $\partial_\zeta f_1(\zeta) = - \text{sign}(\zeta) f_1(\zeta) / a$ and
$\partial_\zeta^2f_1(\zeta) = f_1(\zeta)/a^2-\frac{2}{a}\delta(\zeta)$,
so that
\begin{equation}
I =-\frac{i}{2c_0^2 k_0a}\left(2ik_0a -(k_0a)^2+\frac{(k_0a)^2}{2ik_0a-1}\right).
\label{corrective_term_Keller}
\end{equation}
Thus
\begin{eqnarray}
k^2 &=& k_0^2 +\epsilon^2k_0^2 \left( 3-\frac{4(k_0a)^2}{1+4(k_0a)^2} \right)  \nonumber \\
&& +  2 i \epsilon^2 k_0^2 \left( k_0a+\frac{k_0a}{1+4(k_0a)^2} \right). 
\label{k_Keller}
\end{eqnarray}
This formula gives the damping ${\rm Im}(k)=k_i$ of the coherent wave $\langle \tilde \phi \rangle$ and the effective wave speed ${\omega}/{\rm Re}(k)=c_{\rm eff}$ of the medium. For medium 2, we evaluate the integral numerically.

\section{Keller approximation: Time-dependent random medium \label{Keller_temporal_section}}
Here, we extend the previous analysis to a time-dependent random medium $c(x,t)$. We rewrite the problem as follows:
\begin{equation}
    (L_0+L_1+L_2) \phi = 0,
\end{equation}
where
\begin{align}
    &L_0\phi = -\partial_t^2\phi+c_0^2\partial_x^2(\phi), \\
    &L_1\phi = 2c_0\partial_x^2(\delta c(x,t)\phi), \\
    &L_2\phi = \partial_x^2(\delta c(x,t)^2\phi).
\end{align}
The associated Green's function, solution of $L_0G_0(t,t',x,x')=\delta(t-t')\delta(x-x')$, is
\begin{equation}
    G_0(x,x',t,t') = -\frac{1}{2c_0}\Theta(c_0(t-t')-|x-x'|),
\end{equation}
where $\Theta$ is the Heaviside step function. With these new operators, writing the wave field as
\begin{equation}
    \langle\phi(x,t)\rangle=e^{i(kx-\omega t)},
\end{equation}
it follows that
\begin{align}
&(\langle L_1L_0^{-1}L_1\rangle \langle \phi \rangle) (x,t) \nonumber \\
&= -4c_0^4\epsilon^2 \partial_x^2 \iint_{-\infty}^\infty \mathrm{d}x'\mathrm{d}t' G_0(x,x',t,t') \ \times \nonumber \\
& \qquad \qquad \qquad \qquad \partial_{x'}^2[ f(x'-x)g(t'-t) \langle \phi(t',x') \rangle ]
\label{second_order_term_temporal}
\end{align}
and
\begin{equation}
    \langle L_2 \rangle \langle \phi \rangle = c_0^2\epsilon^2k^2\langle\phi\rangle     .
\end{equation}
The calculations are similar to those for the time-independent random medium. Replacing again $k$ by $k_0$ in the $O(\epsilon^2)$ terms, one gets for medium 1
\begin{equation}
    k^2 = k_0^2\left( 1-\epsilon^2\left[ 1+2c_0\frac{\tau}{a}\frac{1}{Q1}\left( -2+\frac{Q_2}{Q_3}-\frac{Q_4}{Q_5} \right) \right] \right),
\end{equation}
where
\begin{align}
    Q_1 &= 1-i\omega\tau, \\
    Q_2 &= (1-ik_0a)^2, \\
    Q_3 &= 1-ik_0a+Q_1 \frac{a}{\tau c_0}, \\
    Q_4 &= (1+ik_0a)^2, \\
    Q_5 &= -1-ik_0a-Q_1 \frac{a}{\tau c_0}.
\end{align}

We have demonstrated here the possibility to develop a time-dependent theory given the knowledge of the power spectrum (or autocorrelation function) of the perturbation. We note that here too, the solution for medium 2 presented in the corpus is evaluated numerically.

\section{Second-order Born approximation \label{Born_section}}
Another theory is the second-order Born approximation, which we derive here for a time-independent random medium $c(x)$. It is similar to the Keller theory, but one does not look for an effective wave equation satisfied by the mean wave field. Instead, one writes the mean wave field as a series up to a certain order, each term being proportional to a power of $\epsilon$. Using the same notations for the operators as in Appendix \ref{Keller_section}, denoting $\tilde \phi_0$ the unperturbed wave field and $\tilde \phi_1$ the correction such that $\tilde \phi=\tilde \phi_0+\tilde \phi_1$, the 1st-order Born approximation reads
\begin{equation}
\tilde \phi = \tilde \phi_0 -\tilde L_0^{-1}\tilde L_1 \tilde \phi_0+O(\epsilon^2).
\end{equation}
Taking the average, one gets $\langle \tilde \phi \rangle = \tilde \phi_0 + O(\epsilon^2)$. This means that we have to go down to the second order:
\begin{equation}
\tilde \phi = \tilde \phi_0 -\tilde L_0^{-1}\tilde L_1 \tilde \phi_0+\tilde L_0^{-1}\tilde L_1\tilde L_0^{-1}\tilde L_1 \tilde \phi_0-\tilde L_0^{-1}\tilde L_2 \tilde \phi_0+O(\epsilon^3)
\end{equation}
which, averaged, gives 
\begin{equation}
\langle \tilde \phi \rangle = \tilde \phi_0 + \tilde L_0^{-1}\langle \tilde L_1\tilde L_0^{-1}\tilde L_1 \rangle \tilde \phi_0-\tilde L_0^{-1}\langle \tilde L_2 \rangle \tilde \phi_0+O(\epsilon^3).
\end{equation}
We can compute $\langle \tilde L_1 \tilde L_0^{-1}\tilde L_1 \rangle \tilde \phi_0$ and $\langle \tilde L_2 \rangle \tilde \phi_0$ easily because these are mostly equations \ref{second_order_term} and \ref{not_to_forget} replacing $\langle \tilde \phi(x) \rangle = e^{ikx}$ by $\tilde \phi_0(x)=e^{ik_0 x}$. One finally needs to apply $\tilde L_0^{-1}$ which is a convolution by the Green's function. In order to converge, the integration requires a compact support. To model the localization of the perturbation between $X$ and $X+L$, we introduce the window function
\begin{equation}
w(\bar x) = \Theta(\bar x-X)-\Theta(\bar x-(X+L))
\end{equation}
where $\bar x = (x+x')/2$, so that
\begin{equation}
    \langle \delta c(x) \delta c(x')\rangle = \epsilon^2e^{-|\zeta|/a}w(\bar x).
\end{equation}
The approximate solution in $[X,X+L]$ is
\begin{equation}
\langle \tilde \phi(x) \rangle \simeq \tilde \phi_0(x)\left( 1+\epsilon^2\left[ \frac{3}{2}ik_0a-(k_0a)^2+\frac{(k_0a)^2}{2ik_0a-1} \right](x - X) \right),
\end{equation}
which, since $\epsilon \ll 1$, can be written (omitting a phase term) in the form $\langle \tilde \phi(x) \rangle \simeq \textrm{e}^{ik(x-X)}$ where $k$ has the same expression as for the Keller theory (Eq.~\eqref{k_Keller}).
To this level of approximation, the effective $k$ does not depend on $L$.

\section{Spatial homogenization \label{Spatial_homo_section}}
In order to perform the spatial homogenization for a time-independent random medium $c(x)$, we consider the variable
\begin{equation}
\psi = c^2 \phi,
\end{equation}
which is solution of
\begin{equation}
\partial_{t}^2\frac{\psi}{c^2} -\partial_x^2 \psi = 0.
\end{equation}
 Multiplying the equation by $\partial_t \psi$ and integrating over space, then applying an integration by parts, we find that
\begin{equation}
\partial_t \mathcal{E} = 0,
\end{equation}
where
\begin{equation}
\mathcal{E} = \int \mathrm{d}x \left( \frac{1}{2c^2}(\partial_t \psi)^2 + (\partial_x \psi)^2 \right)
\end{equation}
is an expression for the energy. Since it is invariant, we are certain that the homogenization expansion converges.

The medium is assumed to vary on length scales much shorter than the wave (for solar granulation the length scale $a$ is at least shorter than the wave length of acoustic waves). We moreover assume the periodicity of the medium: $c(x)=c(x+a)$. We separate the spatial variable $x$ into $y_0$, a slow-varying spatial scale, and $y_1=y_0/\eta$, a fast-varying spatial scale, where $\eta = k_0(\omega_0) a \ll 1$ \citep[e.g.,][]{2013ApJ...773..101H}. Then
\begin{equation}
\psi = \psi(y_0,y_1,t)
\end{equation}
and
\begin{align}
\partial_x &= \partial_{y_0}+\frac{1}{\eta}\partial_{y_1}, \\
\partial_x^2 &= \partial_{y_0}^2+\frac{2}{\eta}\partial_{y_1}\partial_{y_0}+\frac{1}{\eta^2}\partial_{y_1}^2.
\end{align}
We also expand the solution
\begin{equation}
\psi = \psi_0+\eta \psi_1 + \eta^2 \psi_2 + O(\eta^3),
\end{equation}
where $\psi_i=\psi_i(y_0,y_1,t)=\psi_i(y_0,y_1+a,t)$. We can now proceed to solving the equation order by order.
Order $\eta^{-2}$ gives 
\begin{equation}
\partial_{y_1}^2\psi_0=0.
\end{equation}
Multiplying by $\psi_0$, integrating over $y_1$ and using the argument of periodicity, one gets
\begin{equation}
\int_0^a (\partial_{y_1}\psi_0)^2 \mathrm{d}y_1 = 0,
\end{equation}
meaning that $\psi_0$ does not depend on $y_1$. 
Order $\eta^{-1}$ then gives 
\begin{equation}
\partial_{y_1}^2 \psi_1=0,
\end{equation}
meaning that $\psi_1$ does not depend on $y_1$ either. 
Finally, at order $\eta^0$,
\begin{equation}
\partial_t^2 \frac{\psi_0}{c^2} - \partial_{y_0}^2 \psi_0 - \partial_{y_1}^2 \psi_2 = 0.
\end{equation}
Integrating over the fast-varying coordinate $y_1$, invoking periodicity, one finds the following homogenized equation for $\psi$:
\begin{equation}
\partial_t^2 \psi_0 - \frac{1}{\overline{c^{-2}}} \partial_{y_0}^2 \psi_0 = 0,
\end{equation}
where $\overline{c^{-2}}=\frac{1}{a}\int_0^a c^{-2}\mathrm{d}y_1$ is a spatial average. The homogenization method, used here for a periodic medium, has been generalized to a statistically homogeneous and ergodic random medium, by making the period tend to $\infty$ \citep[e.g.,][]{Papanicolaou1982}. The spatial average identifies then with the statistical average. The homogenized sound speed $c_{\rm h}$ of the medium is therefore equal to $\langle c^{-2} \rangle^{-1/2}$. Knowing that $c = c_0+\delta c$, $c^{-2}\simeq c_0^{-2}(1-2\delta c/c_0+3 \delta c^2/c_0^2)$ and $\langle c^{-2} \rangle \simeq c_0^{-2}(1+3\epsilon^2)$. Hence:
\begin{equation}
c_{\rm h}=\langle c^{-2} \rangle^{-1/2} \simeq c_0(1-\frac{3}{2}\epsilon^2).
\end{equation}
We note that the spatial homogenization technique does not make an attenuation arise.

\section{Ray approximation \label{Geom_optics_section}}
The geometrical optics theory, or ray theory, is an infinite-frequency approximation. In practice the applicability conditions are \citep{1989psr4.book.....R}:
\begin{align}
    \epsilon &\ll 1, \\
    k_0a & \gg 1, \\
    k_0a & \gg 2\pi \frac{L}{a}.
\end{align}
Under these conditions, the wave travel time inside the random medium starting at $x=X$ is computed as an integral of the slowness over the ray path:
\begin{equation}
    t = \int_{X}^x c^{-1}(s)\ \mathrm{d}s
    = (x-X)\ \overline{  c^{-1} }  .
\end{equation}
Assuming ergodicity of the random medium, the spatial average identifies with the statistical average and
\begin{equation}
    c_{\rm ray} = \langle c^{-1} \rangle ^{-1} .
\end{equation}

\section{Comparing theories with numerical simulations in the limit \texorpdfstring{$\epsilon\rightarrow 0$}{epsilon tends to zero} \label{Comparison_theories_section}}

\begin{figure*}[t]
\centering
\includegraphics[width=\textwidth]{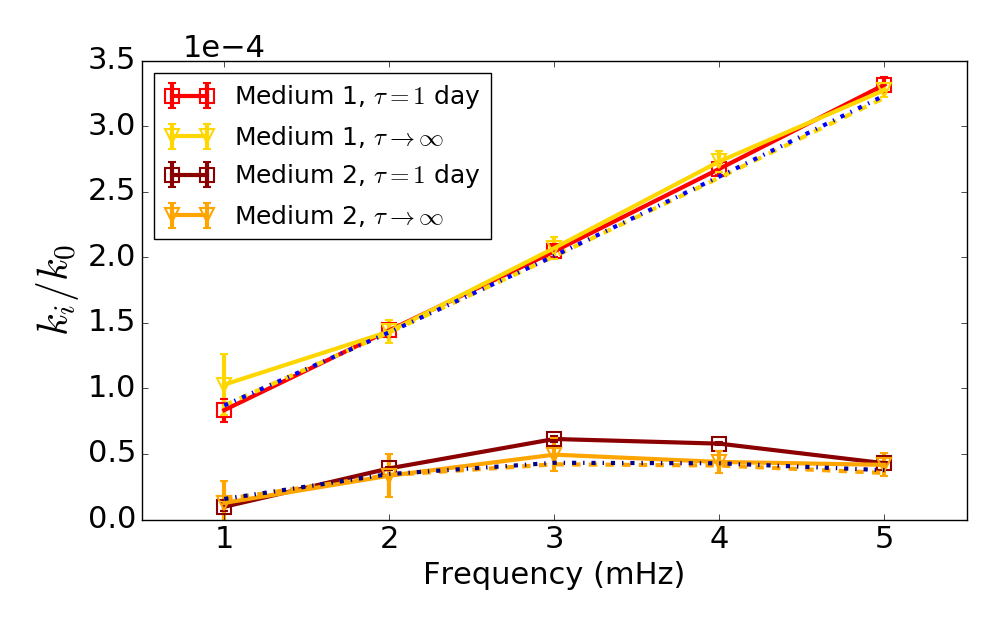}
\includegraphics[width=\textwidth]{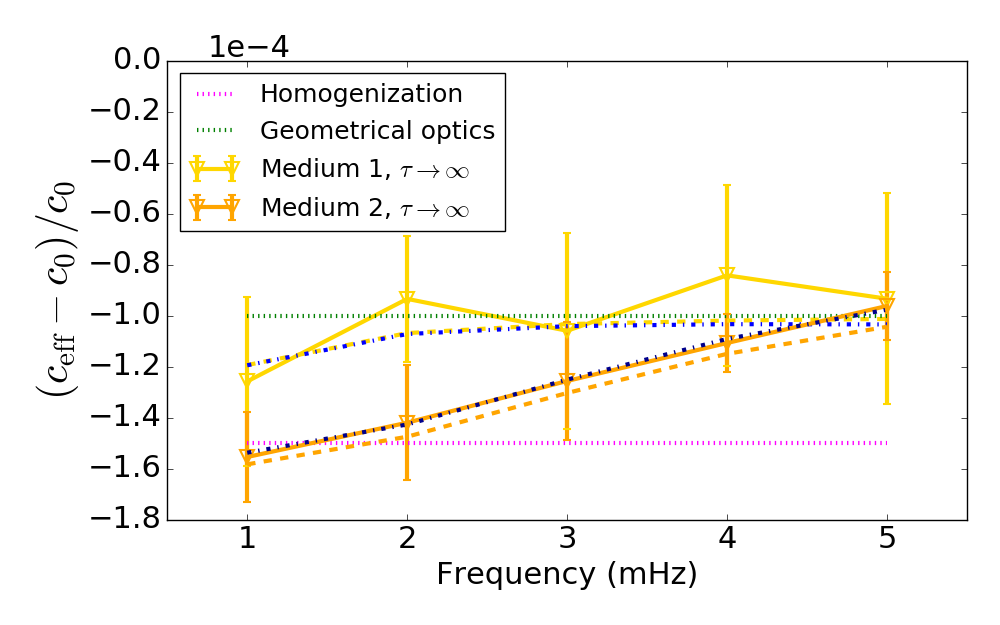}
\caption{Comparison of theories with simulations for the average wave field ($\epsilon=0.01$). Top: attenuation. Bottom: effective wave speed. The triangles are for the simulations in frequency domain ($\tau \rightarrow \infty$), the squares for those in time domain ($\tau=1$~day). The two dashed-dotted blue lines are the Born solutions for media 1 and 2, while the yellow and orange dashed lines are the Keller solutions. 1-$\sigma$ error bars are shown.}
\label{comparison_theories_simu}
\end{figure*}

Fig.~\ref{comparison_theories_simu} summarizes the accuracy of the (frozen) Keller theory, the Born second-order approximation, the spatial homogenization and the ray theory in the small-perturbation regime ($\epsilon=0.01$). For each simulation, ten sets of $10^4$ realizations were generated to get the error bars. For such a small perturbation, we are in the regime of validity of the Born and Keller theories and the results are in agreement with the numerical simulations for the attenuation and the effective wave speed. The attenuation for medium 2 resulting from the time-domain simulation differs from the attenuation from the frequency-domain one, likely because of numerical diffusion. As $k_0 a$ is of order unity in our setup, we are not a priori in the regime of validity of the homogenization or the geometrical optic theories. However, the geometrical optics is in good agreement with the numerical simulations for medium 1, despite the fact that the condition $k_0a \gg 2\pi \frac{L}{a}$ is not verified in our simulations. Medium 2 exhibits, just like for $\epsilon=0.1$, a transition from the homogenization regime at small frequencies ($<1$~mHz) to the geometrical optics regime at high frequencies ($>5$~mHz).

\end{appendix}

\end{document}